\documentclass[aps,prb,preprint,showpacs,superscriptaddress,floatfix]{revtex4}
\usepackage{graphics}
\usepackage{epsfig}
\usepackage{amsmath}
\usepackage{amssymb}
\usepackage{calc}
\newcommand{\beq}{\begin{equation}}
\newcommand{\eeq}{\end{equation}}
\newcommand{\bea}{\begin{eqnarray}}
\newcommand{\eea}{\end{eqnarray}}

\begin{document}
\title{Hitchin functionals are related to measures of entanglement}
\author{P\'eter L\'evay and G\'abor S\'arosi}
\affiliation{Department of Theoretical Physics, Institute of
Physics, Budapest University of Technology, H-1521 Budapest,
Hungary}
\date{\today}
\begin{abstract}
According to the Black Hole/Qubit Correspondence (BHQC) certain black hole entropy
formulas in supergravity can be related to  multipartite entanglement
measures of quantum information.
Here we show that the origin of this correspondence is
a connection between Hitchin functionals used as action functionals
 for form theories of gravity
 related to topological strings, and entanglement measures for systems
 with a small number of constituents.
The basic idea acting as a unifying agent in these seemingly
unrelated fields is stability connected to the mathematical notion
of special prehomogeneous vector spaces associated to Freudenthal
systems coming from simple Jordan algebras. It is shown that the
nonlinear function featuring these functionals and defining
Calabi-Yau and generalized Calabi-Yau structures is the
Freudenthal dual a concept introduced recently in connection with
the BHQC. 
We propose to use the Hitchin invariant for three-forms
in $7$ dimensions as an entanglement measure playing a basic role
in classifying three-fermion systems with seven modes. The
representative of the class of maximal tripartite entanglement is
the three-form used as a calibration for compactification on
manifolds with $G_2$ holonomy. The idea that entanglement measures
are related to action functionals from which the usual
correspondence of the BHQC follows at the tree level suggests that
one can use the BHQC in a more general context.

\end{abstract}
 \maketitle{}

\renewcommand\thesection{\arabic{section}}
\renewcommand\thesubsection{\arabic{subsection}}

\numberwithin{subsection}{section}
\numberwithin{subsubsection}{subsection}

\section{Introduction}

The main motivation of the present paper is to generalize further the recently  discovered Black Hole/Qubit Correspondence (BHQC)\cite{review,review2}.
This correspondence is based on striking mathematical connections found recently between
two seemingly unrelated research areas:
black hole solutions in String Theory\cite{Becker} (ST) and
the theory of multipartite entanglement measures\cite{Plenio}
in Quantum Information (QI)\cite{Nielsen}.

The main correspondence is between the structure of the Bekenstein-Hawking
entropy formulas in extremal BPS or non BPS black hole solutions
in supergravity and certain multipartite entanglement measures of
composite quantum systems with either distinguishable or indistinguishable
constituents\cite{Duff1,Kalloshlinde,Levay1,Duff2,Levayfano,Duff3}.
As an other aspect of the correspondence it has also been realized that
the classification problem of entanglement types of special entangled systems
and special types of black hole solutions can be mapped to each other\cite{review,Kalloshlinde,Borsten4qbit1}.
Using this input coming from the physics of black holes the BHQC motivated the introduction of new entanglement measures\cite{levvran1,levvran2} and helped to classify the entanglement patterns of certain quantum systems with a small number of constituents\cite{levvran1,levvran2,Borsten4qbit1,Borsten3qbit}.

Apart from structural correspondences between measures and classes of entanglement and entropy formulas and classes of solutions for black holes, the BHQC also addressed issues of dynamics.
In particular the dynamics of moduli stabilization related to the attractor mechanism\cite{FKS} has been shown to correspond to a distillation procedure of entangled states of very special kind on the black hole event horizon\cite{Levay1,Levay3}.
This result has recently been demonstrated in the IIB picture via the use of entangled states associated to wrapping configurations of three-branes\cite{wrapornottowrap,qfromextra} on $T^6$. These complex states are depending on the black hole charges and the complex structure moduli\cite{qfromextra}.

The BHQC has revealed the interesting finite geometric structure of black hole entropy formulas\cite{hexa,cherchiai} and related them to Mermin squares\cite{Mermin}, error correcting codes\cite{review} and graph states\cite{Levay3} objects playing
an important role in quantum information.
It is also important to note that algebraic structures like Freudenthal triple
systems\cite{Freudenthal,McCrimmon,Krutelevich} that has already been
well-known to the supergravity community\cite{GST1}
has made their debut to the theory of quantum entanglement via the BHQC
\cite{levvran1,Borstene7,levvran2}.
On the other hand reconsidered in the
light of quantum information applications of these
systems to the physics of black holes have also resulted in
introducing useful notions such as black holes admitting a
Freudenthal and Jordan dual\cite{freuddual}.

As the main reason for the BHQC usually the occurrence of similar
{\it symmetry structures} is emphasized\cite{review,review2}. Indeed, on
the string theory side there are the U-duality groups\cite{Hull}
leaving invariant the black hole entropy formulas, on the other
hand on the quantum information theoretic side there are the
groups of admissible transformations\cite{Bennett} used to
represent local manipulations on the entangled subsystems leaving
invariant the corresponding entanglement measures. The U-duality
groups in ST are real but the groups of admissible transformations
in QI are complex. However, under special circumstances the real
U-duality groups should be embedded into the complex domain using
reality
conditions\cite{Levay1,Borsten4qbit1,Levay4q,qfromextra}
rendering the techniques of QI applicable.

In this paper as another reason for the BHQC we would like to
propose the notion of {\it stability}. This notion turns out to be
a useful one since via the attractor mechanism it can naturally be
related to the dynamic aspects of the BHQC. Interestingly in the
separated research areas of string and quantum information theory
the idea of stability appeared nearly at the same time. In quantum
information Klyachko proposed\cite{Klyachko} (semi)stability as a
useful idea to capture entanglement for a system characterized by
a dynamical symmetry group. In this context the role of special
invariants as entanglement measures separating unstable orbits
from stable ones has been emphasized. On the string theory side
research was initiated by the influential mathematical papers of
Hitchin on stable forms\cite{stable} and their connection to
generalized Calabi-Yau spaces\cite{Hitchin,genHitchin} and
manifolds with special holonomy. In these papers functionals based
on nondegenerate stable forms have been constructed. It is then
shown that their extremal properties are related to the existence
of special geometric structures on six, seven and eight
dimensional manifolds. The possibility for introducing these
structures rests on the applicability of an important notion. This
idea is well-known to mathematicians however, not yet fully
appreciated by physicists. This is the notion of a {\it
prehomogeneous vector space} (PV) as introduced by Sato and Kimura
in their classical work\cite{Satokimura,Kimura}.

A prehomogeneous vector space is a triple $(G,R,V)$ where $V$
is a finite dimensional vector space over ${\mathbb C}$, $G$ is a
linear algebraic group and $R$ is a rational representation
$R:G\to GL(V)$ such that
 for a generic element $v\in V$ $G$ has an open dense orbit $\rho(G)v$ in $V$.
An element $v\in V$  is called {\it stable} if it lies in such an open orbit of $G$.

In the setting of entanglement $V$ should correspond to the state
space of our quantum system consisting of a finite number of
subsystems. In the case of distinguishable constituents $V$ has a
tensor product structure. For indistinguishable ones we have
either the symmetric or the antisymmetric tensor product structure
corresponding to bosons or fermions. The group action $G$ and its
orbit should represent the admissible local operations on the
subsystems and the {\it generic} entanglement class respectively.
One is then left to define polynomial invariants called
entanglement measures, which are usually relative invariants. This
means that they are invariants up to a character of $G$. In this
picture the open dense orbit should be characterized by the {\it
nonvanishing} of a particular relative invariant. Stability then
would mean that states in a neighborhood of a particular one are
equivalent with respect to the group $G$ of local manipulations.

Clearly using stability for a definition of entanglement via PVs
is too restrictive. (That was the reason for using the notion of
semistability instead). This is because for prehomogeneous vector
spaces one should have\cite{Satokimura} ${\rm dim}G-{\rm
dim}G_v={\rm dim}V$ where $G_v$ is the stabilizer of a $v\in V$.
Since the dimension of $G$ has a slower growth than the dimension
of $V$ the notion of stability as related to PVs works only for
characterizing {special entangled systems}. These are the ones
with a small number of constituents. Remarkably such systems are
the ones also related to the BHQC.

On the string theory side in form theories of quantum
gravity\cite{Dijkgraaf} PVs show up via the use of {\it stable
forms}\cite{stable}. In $6$ dimensions these form theories are
related to topological strings\cite{topstring}. In $7$ dimensions
they define the low energy limit of topological $M$-theory.
Generally these form theories are based on action principles for
$p$-forms on a manifold $M$. The actions involve a volume element
constructed in a nontrivial way from the $p$-form. At each point
of the manifold $M$ the vector space $V$ for the PV is arising as
the space $\wedge^pW^{\ast}$ where $W$ is the  tangent space at a
point of $M$. Now in this context a $p$-form $\varrho$ is stable at
$m\in M$ if it lies in an open orbit of the $GL(W)$ action on
$\wedge^pW^{\ast}$. $\varrho$ is stable if it is stable at every
point of $M$.

The physical significance of stable forms stems from the OSV
conjecture\cite{OSV}. OSV suggested a relation between black hole
entropy and the partition function of topological strings. Later
work have revealed\cite{Nekrasov,Dijkgraaf} that at the classical
level black hole entropy and the topological string partition
function are also related to Hitchin's functional\cite{Hitchin} for
real three-forms on a $6$ dimensional manifold $M$. The critical
points of Hitchin's functional define a Calabi-Yau structure on
$M$. Then the idea was to use Hitchin's functional also to recover
the quantum corrections that has already been calculated via
topological string techniques\cite{Cecotti}. It turned
out\cite{PW} that in order to achieve agreement at one loop level one
has to use the so called generalized Hitchin
functional\cite{genHitchin} instead. Now this new generalized
functional\cite{genHitchin} contains
 polyforms of either even or odd degree, with its critical points
 defining
 generalized Calabi-Yau spaces\cite{genHitchin,Gualtieri}.
A further generalization occurs if we are considering form theories of gravity in $7$ dimensions.
Here the PVs in question are based on the vector spaces $\wedge^3\tilde{W}^{\ast}$ and
$\wedge^4\tilde{W}^{\ast}$ where now the seven dimensional vector space $\tilde{W}$ is the tangent space of the $7$-manifold ${\tilde M}$ at a point. The group $G$ is $GL(\tilde{W})$ and the stabilizer
of a generic form is the exceptional group $G_2$.
Using these stable forms then one can define functionals that via their critical points generating $G_2$
holonomy on the $7$-manifold.
There is a natural connection\cite{stable} between these functionals and the Hitchin functionals of the $6$-manifold $M$ with critical points being manifolds with $SU(3)$ holonomy.
This connection gives rise to a relation\cite{Dijkgraaf} between topological $M$-theory on the $7$-manifold $\tilde{M}$ and topological string theory on the $6$-manifold $M$.

The proposal we would like to put forward in this paper is to regard
the invariants  underlying these functionals as entanglement measures for special entangled systems with the class of stable forms corresponding to the class of genuine entangled states.
This idea  makes it possible to generalize the BHQC substantially.
First of all entanglement measures are now related to {\it action functionals}
from which one can recover at the {\it semiclassical level} the usual correspondence of the BHQC found between the Bekenstein Hawking entropy and some entanglement measure.
However, since one loop calculations\cite{PW} based on {\it quantization} of such functionals are also capable of reproducing results obtained
by topological string techniques\cite{Cecotti},
this interpretation also suggests that one can use the entanglement measures of the BHQC in a more general context.
Secondly since
posessing a stable form is far less restrictive than the
requirement of special holonomy, in this generalized version of the BHQC one does not have to assume the metric to be of the special holonomy (Calabi-Yau etc.) form.
Thirdly,
after identifying Hitchins invariants with entanglement measures
a reconsideration of the results of the BHQC on the attractor mechanism\cite{Levay1,Levay3} provides a new way of looking at the dynamic aspects of the BHQC.

The organization of the paper is as follows. In Section 2. we
summarize the basic material concerning the simplest of tripartite
entanged systems both for distinguishable and indistinguishable
constituents. In Section 3. we introduce Hitchin's functional for
the real three form $\varrho$ with the underlying invariant
related to the canonical entanglement measure for three fermions
with six single particle states. Here as a novelty the usual
nonlinear function $\hat{\varrho}(\varrho)$ is expressed in terms
of the Freudenthal dual\cite{freuddual} originating from the cubic
Jordan algebra of $3\times 3$ complex matrices\cite{Krutelevich}.
In terms of the associated Freudenthal system the symplectic
structure, and corresponding the Hamiltonian system\cite{Hitchin}
is expressed in an elegant manner. In special subsections of
Section 3. we also consider truncations giving rise to entangled
systems with $SP(6,{\mathbb C})$  and $SP(2,{\mathbb C})^{\times
3}$ as the group of admissible transformations. Stable forms of
the corresponding real cases gives rise to familiar structures
known from type IIB compactifications on $T^6$ and $T^2\times
T^2\times T^2$ (STU-model). A further truncation with three
bosonic qubits corresponds to the $t^3$ model. In Section 4. we
consider the generalized Hitchin functional. It is shown that the
corresponding invariant gives rise to an entanglement measure for
a fermionic system with six single particle states with either an
even or add number of particles. An alternative interpretation can
also be given in terms of the tripartite entanglement of six
qubits a structure living naturally inside the recently discovered
tripartite entanglement of seven
qubits\cite{Duff2,Levayfano,cherchiai}. Instead of the usual way
of writing this invariant in terms of pure spinors we present a
Freudenthal triple based description which is coming from the
cubic Jordan algebra of  $3\times 3$ matrices with biquaternionic
entries. As an illustration on the string theory side we relate
this invariant to the work of Pestun for $N=2$ compactification on
$T^6$ with more general backgrounds. Section 5. we propose to use
Hitchins invariant based on three-forms in $7$ dimensions as an
entanglement measure playing a basic role in classifying
three-fermion systems with seven single particle states. We
reinterpret the classification of three-forms in $7$
dimensions\cite{Cohen} as the classification of entanglement
classes under the so called SLOCC group\cite{Bennett} of quantum
information. We emphasize that the representative of the class of
maximal tripartite entanglement is related to the usual three-form
used as a calibration for compactifications on manifolds with
$G_2$ holonomy and the structure of the octonions. Section 6. is
left for the comments and the conclusions. Here we also speculate
on the meaning of our entangled "states", and suggest to try to
connect them via the OSV conjecture to topological string theory.
These attempts might pave the way for finding the physical basis
of the BHQC. For the convenience of the reader in an Appendix
(Section 8.) we summarized the material needed for a Freudenthal
triple based description of the generalized Hitchin functional.

\section{Entanglement}

\subsection{Distinguishable constituents}
In Quantum Information Theory (QI) entangled systems with
distinguishable constituents are represented by vectors in a tensor
product of finite dimensional Hilbert spaces\cite{Nielsen,Plenio}. In the special case
of pure states of a multiqubit system states are elements of the
complex vector space ${\mathbb C}^2\otimes {\mathbb
C}^2\otimes\dots \otimes { \mathbb C}^2$ where the number of
two-state spaces equals the number of qubits. For example a
three-qubit state can be written in the form \beq \vert
\psi\rangle =\sum_{i,j,k=0,1}\psi_{ijk}\vert
i\rangle_1\otimes\vert j\rangle_2\otimes\vert k\rangle_3\in
{\mathbb C}^2\otimes{\mathbb C}^2\otimes {\mathbb C}^2 \eeq
\noindent where the subscripts $1,2,3$ refer to the
distinguishable subsystems. Since in QI entanglement is regarded
as a resource for performing different tasks, the central problem
is to characterize different types of entanglement. Since
entanglement is a global phenomenon, local transformations are
supposed to have no effect on the entanglement types. According to
this idea entanglement types of say three qubits should correspond
to different {\it orbits} under some set of local transformations
of the form \beq \vert \psi\rangle \mapsto (S_1\otimes S_2\otimes
S_3)\vert\psi\rangle. \eeq Here a specification of the local
operators $S_1,S_2,S_3$ defines a classification scheme of
entanglement types. The restriction for these operators to be unitary is an
obvious choice however, for practical reasons other classification
schemes proved to be useful. Chosing local equivalence under
the action of operators belonging to the group $GL(2,{\mathbb C})$
yields the so called SLOCC orbits\cite{Bennett,Dur}. The name comes from the
abbreviation of Stochastic Local Operations and Classical
Communication, referring to the particular type of protocols that
can mathematically be represented by such invertible complex
linear operators.

Entanglement measures are certain polynomials in the amplitudes of
$\vert\psi\rangle$ satisfying a number of physically sensible
criteria. For our concern the most important of these criteria is
that they should be (relative) invariants under the action of the
SLOCC group. In our special case of three qubits the quartic
polynomial\cite{Cayley,Kundu,Gelfand}

\begin{eqnarray}
D(\psi)&=&
[\psi_0\psi_7-\psi_1\psi_6-\psi_2\psi_5-\psi_3\psi_4]^2-
4[(\psi_1\psi_6)(\psi_2\psi_5)+(\psi_2\psi_5)(\psi_3\psi_4)\\\nonumber&+&
(\psi_3\psi_4)(\psi_1\psi_6)]+
4\psi_1\psi_2\psi_4\psi_7+4\psi_0\psi_3\psi_5\psi_6 \label{Cayley}
\end{eqnarray}
where $(\psi_0,\psi_1,\dots,\psi_7)\equiv
(\psi_{000},\psi_{001},\dots,\psi_{111})$, gives rise to a famous
entanglement measure called the {\it three-tangle}\cite{Kundu} which for normalized states satisfies \beq 0\leq
{\tau}_{123}=4\vert D(\psi)\vert\leq 1. \label{threetangle} \eeq
\noindent Under SLOCC transformations $D(\psi)$ transforms as \beq
D(\psi)\mapsto ({\rm Det}S_1)^2 ({\rm Det}S_2)^2({\rm
Det}S_3)^2D(\psi) \eeq \noindent hence this polynomial is a
relative invariant.

 The classification problem of SLOCC
 entanglement types has been solved by mathematicians\cite{Gelfand}, the proof
 has later been independently rediscovered by physicists\cite{Dur}. According
 to this result there are six nontrivial SLOCC entanglement classes.
 The genuine entanglement class with normalized representative is the so called GHZ-class\cite{GHZ}
\beq \vert GHZ\rangle=
\frac{1}{\sqrt{2}}(\vert 000\rangle +\vert 111\rangle) \label{GHZ}
\eeq \noindent It is characterized by the constraint $D(\psi)\neq 0$.
The so called  W-class\cite{Dur} represented by \beq \vert
W\rangle=\frac{1}{\sqrt{3}}(\vert 001\rangle+\vert
010\rangle+\vert 100\rangle) \label{W} \eeq \noindent has
$D(\psi)=0$ however, sates belonging to this class still contain
some sort of tripartite entanglement\cite{Dur}. The remaining four classes
are separable. This means that their representatives are either of
the form $\frac{1}{\sqrt{2}}(\vert 00\rangle +\vert
11\rangle)\otimes \vert 0\rangle$ or two similar states with the
qubits cyclically permuted (biseparable states), or represented by
$\vert 000\rangle$ (totally separable states).

The corresonding classification of SLOCC entanglement types over
the reals (i.e. the classification for three rebits\cite{Caves}) has also
been used by physicists\cite{Acin}. In this case rebits live in ${\mathbb
R}^2\otimes {\mathbb R}^2\otimes {\mathbb R}^2$, and the SLOCC
group is three copies of $GL(2,{\mathbb R})$. The result in this
case is that the GHZ-class splits into two classes. One of them is
the usual one with representative as given by Eq.(\ref{GHZ}), and
$D(\psi)>0$. However, now we have an extra class with $D(\psi)<0$
with representative \beq \vert GHZ\rangle_-=\frac{1}{2}(\vert
000\rangle-\vert 011\rangle-\vert 101\rangle -\vert 110\rangle).
\label{GHZ2} \eeq\noindent Note that the state \beq \vert
GHZ\rangle_+=\frac{1}{2}(\vert 000\rangle+\vert 011\rangle+\vert
101\rangle +\vert 110\rangle) \label{GHZ1} \eeq \noindent with
$D(\psi)>0$ is real SLOCC equivalent to the one of Eq.(\ref{GHZ}).
Indeed \beq \vert GHZ\rangle =(H\otimes H\otimes H)\vert
GHZ\rangle_{+},\qquad
H=\frac{1}{\sqrt{2}}\begin{pmatrix}1&1\\1&-1\end{pmatrix} \eeq
\noindent where $H$ is the Hadamard matrix of discrete Fourier
transformation.
Notice also that the new state $\vert GHZ\rangle_-$ as a real one can be embedded into ${\mathbb C}^2\otimes {\mathbb C}^2\otimes {\mathbb C}^2$ as a state
\beq
\vert GHZ\rangle_-=\frac{1}{
\sqrt{2}}(\vert\alpha\rangle\otimes\vert\alpha\rangle\otimes\vert\alpha\rangle+
\overline{{\vert\alpha\rangle}\otimes
{\vert\alpha\rangle}\otimes{\vert\alpha\rangle}}),
\qquad \vert\alpha\rangle=\frac{1}{\sqrt{2}}(\vert 0\rangle +i\vert 1\rangle).
\label{ketto}
\eeq
\noindent

\subsection{Indistinguishable constituents. Fermions}

The notion of entanglement can also be generalized to include systems with
indistinguishable parts\cite{Schlie}.
In the following we need results merely from the theory of fermionic entanglement.
We consider fermions on an $M$ dimensional single particle Hilbert space
$V={\mathbb C}^M$. The observables are generated by the operators $f^{\dagger}$
and $f$ satisfying the usual canonical anticommutation relations $\{f_k,{f_l}^{\dagger}\}={\delta}_{kl}$, $\{f_k,f_l\}=0$
$\{{f_k}^{\dagger},{f_l}^{\dagger}\}=0$
.
It is clear that $f_j^{\dagger}$ creates a particle in the mode, or single particle state, corresponding to the basis vector $e_j$ of ${\mathbb C}^M$.
The Hilbert space of the fermionic system is spanned by the basis
\beq
\left(f_1^{\dagger}\right)^{n_1}
\left(f_2^{\dagger}\right)^{n_2}\dots \left(f_M^{\dagger}\right)^{n_M}\vert 0\rangle
\eeq
\noindent
where $n_j\in\{0,1\}$ and the vacuum state $\vert 0\rangle$ satisfies $f_j\vert 0\rangle =0, \forall j$.
The $N$ particle subspace of the Fock space is spanned by those vectors that satisfy the constraint $\sum_j n_j=N$.

For example an (unnormalized) three fermion state ($N=3$) with six single particle states or {\it modes} ($M=6$) is represented by the state vector
\beq
\vert P\rangle =\sum_{1\leq i_1<i_2<i_3\leq 6}P_{i_1i_2i_3}f_{i_1}^{\dagger}
f_{i_2}^{\dagger}f_{i_3}^{\dagger}\vert 0\rangle
\label{itt1}
\eeq
\noindent
where the $P_{i_1i_2i_3}$ are $20$ complex amplitudes characterizing the state.
It is convenient to use another representation for such fermion states as multilinear forms. Hence the state $\vert P\rangle$ can also be represented by a {\it three-form} over the space $V={\mathbb C}^6$ as
\beq
P=\sum_{1\leq i_1<i_2<i_3\leq 6}P_{i_1i_2i_3}e^{i_1}\wedge e^{i_2}\wedge e^{i_3}\in \wedge^3V^{\ast}
\eeq
\noindent
where $\{e^j\}, j=1,\dots 6$ are basis vectors of $V^{\ast}$ dual to the
basis vectors $\{e_j\}$ of $V$.

Due the indistinguishable nature of the subsystems SLOCC transformations are acting on our fermion states with the {\it same} $GL(M,{\mathbb C})$
transformations to be applied to each slot. For example
for $V={\mathbb C}^6$
the  SLOCC transformation
$\vert P\rangle \mapsto (S\otimes S\otimes S)\vert P\rangle$ is represented by
\beq
P_{i_1i_2i_3}\mapsto P_{j_1j_2j_3}{S^{j_1}}_{i_1}
{S^{j_2}}_{i_2}{S^{j_3}}_{i_3}, \qquad S={S^j}_ie^j\otimes e_i\in GL(V)
\eeq
\noindent
coming from  the transformation rule $P\mapsto S^{\ast}P$ for three-forms.

There is a quartic polynomial which is a relative invariant with respect to the SLOCC group\cite{Satokimura,levvran1}.
In order to define this polynomial
we reorganize the $20$ independent complex amplitudes $P_{i_1i_2i_3}$ into two
complex numbers $\eta,\xi$ and two complex $3\times 3$
matrices $X$ and $Y$ as follows. As a first step we change our
labelling convention by using the symbols
$\overline{1},\overline{2},\overline{3}$ instead of $4,5,6$
respectively. The meaning of the labels $1,2,3$ is not changed.
Hence for example we can alternatively refer to $P_{456}$ as
$P_{\overline{1}\overline{2}\overline{3}}$ or to $P_{125}$ as
$P_{12\overline{2}}$. Now we define
\beq
\eta\equiv P_{123},\qquad \xi\equiv P_{\overline{123}}
\label{etaxi}
\eeq
\beq
X=\begin{pmatrix}X_{11}&X_{12}&X_{13}\\X_{21}&X_{22}&X_{23}\\X_{31}&X_{32}&X_{33}\end{pmatrix}
\equiv\begin{pmatrix}P_{1\overline{23}}&P_{1\overline{31}}&P_{1\overline{12}}\\
P_{2\overline{23}}&P_{2\overline{31}}&P_{2\overline{12}}\\P_{3\overline{23}}&P_{3\overline{31}}&P_{3\overline{12}}\end{pmatrix},
\label{ymatr}
\eeq
\beq
Y=\begin{pmatrix}Y_{11}&Y_{12}&Y_{13}\\Y_{21}&Y_{22}&Y_{23}\\Y_{31}&Y_{32}&Y_{33}\end{pmatrix}\equiv
\begin{pmatrix}P_{\overline{1}23}&P_{\overline{1}31}&P_{\overline{1}12}\\P_{\overline{2}23}&P_{\overline{2}31}&P_{\overline{2}12}\\
P_{\overline{3}23}&P_{\overline{3}31}&P_{\overline{3}12}\end{pmatrix}.
\label{Xmatr}
\eeq
With this notation the quartic polynomial is
\beq
{\cal D}(P)=[\eta\xi -{\rm
 Tr}(XY)]^2-4{\rm Tr}(X^{\sharp}Y^{\sharp})+4\eta{\rm
 Det}(X)+4\xi{\rm Det}(Y) \label{Cayleygen}
 \eeq
 \noindent where $X^{\sharp}$ and $Y^{\sharp}$ correspond to the
 regular adjoint matrices for $X$ and $Y$ hence for example
 $XX^{\sharp}=X^{\sharp}X={\rm Det }(X)I$
 with $I$ the $3\times 3$ identity matrix (see also Eq.(\ref{sharp}) in the Appendix).
Clearly the structure of our new polynomial ${\cal D}(P)$ is very
similar to the one of $D(\psi)$ we defined in Eq.(\ref{Cayley}).
Later on this will be important for us. ${\cal D}(P)$ defines an
entanglement measure similar to the three-tangle in the
form\cite{levvran1} \beq 0\leq {\cal T}_{123}=4\vert{\cal
D}(P)\vert \label{tanglegen} \eeq \noindent where ${\cal
T}_{123}\leq 1$ for normalized states.

There is an alternative way of describing this polynomial. Let us
define the a symplectic form on $\wedge^3V^{\ast}$ as follows \beq
\{\cdot,\cdot\}:\wedge^3V^{\ast}\times\wedge^3V^{\ast}\to {\mathbb
C},\qquad (P,Q)\mapsto
\frac{1}{3!3!}{\varepsilon}^{abcijk}P_{abc}Q_{ijk} \label{sympl}
\eeq \noindent where the $2\times 20$ amplitudes of $P$ and $Q$
has been extended to totally antisymmetric tensors of rank three
and we used the summation convention. Now we define 
$\tilde{P}$ for the three-form $P\in \wedge^3V^{\ast}$ as \beq
\tilde{P}=\frac{1}{3!}\tilde{P}_{abc}e^a\wedge e^b\wedge e^c,\quad
\tilde{P}_{abc}=\frac{1}{2!3!}{\varepsilon}^{di_2i_3i_4i_5i_6}P_{bcd}P_{ai_2i_3}P_{i_4i_5i_6}.
\label{dual} \eeq \noindent The quartic invariant then takes the
form \beq {\cal D}(P)=\frac{1}{2}\{\tilde{P},P\}.
\label{quarticinvper2} \eeq \noindent In the theory of Freudenthal
triple systems the quantity $\tilde{P}$ which is cubic in the
original amplitudes of $P$ is usually defined via the so called
trilinear form\cite{Krutelevich}. With the help of $\tilde{P}$ for
a state with ${\cal D}\neq 0$ one can define the quantity \beq
\hat{P}\equiv \frac{-\tilde{P}}{\sqrt{\vert {\cal D}\vert}}.
\label{freudual} \eeq \noindent $\hat{P}$ is the {\it Freudenthal
dual} of $P$ as defined by the paper\cite{freuddual} of Borsten
et.al.

The classification problem for three-forms in $V={\mathbb C}^6$
has been solved long ago by mathematicians\cite{Reichel}, in the
context of fermionic entanglement it has recently been
rediscovered by physicists \cite{levvran1}. According to this
result, we have four disjoint SLOCC classes. The representatives
of these classes can be brought to the following form

\begin{equation}
P=\frac{1}{2}(e^1\wedge e^2\wedge e^3+e^1\wedge
e^{\overline{2}}\wedge e^{\overline{3}}+e^2\wedge
e^{\overline{3}}\wedge e^{\overline{1}}+e^3\wedge
e^{\overline{1}}\wedge e^{\overline{2}}),\quad {\cal
D}(P)\neq 0 \label{1}
\end{equation}
\noindent
\begin{equation}
P=\frac{1}{\sqrt{3}}(e^1\wedge e^2\wedge e^3+e^1\wedge
e^{\overline{2}}\wedge e^{\overline{3}}+e^2\wedge
e^{\overline{3}}\wedge e^{\overline{1}}),\qquad {\cal
D}(P)=0,\quad \tilde P\neq 0
\end{equation}
\noindent
\begin{equation}
P=\frac{1}{\sqrt{2}}e^1\wedge (e^2\wedge e^3+
e^{\overline{2}}\wedge e^{\overline{3}}),\qquad {\cal
D}(P)=0,\quad \tilde{P}=0
\end{equation}
\noindent
\begin{equation}
P=e^1\wedge e^2\wedge e^3,\qquad {\cal D}(P)=0,\quad
\tilde{P}=0. \label{4}
\end{equation}
\noindent
In analogy with the three-qubit case we will refer to the first two classes as the GHZ and W-class.
In order to separate the last two classes (i.e. the biseparable and separable ones) one has to use the Pl\"ucker relations\cite{levvran1}.
Clearly the GHZ and W classes are the two inequivalent classes for tripartite entangled fermionic systems with six modes. These classes are completely characterized by the relative invariant ${\cal D}(P)$ and the dual state $\tilde{P}$ (a covariant).
The GHZ class corresponds to a stable orbit. This fact is related to the result that our system corresponds to a PV which is the class No.5. in the Sato-Kimura classification\cite{Satokimura}.
As in the previous section after restricting to the real case the GHZ class splits into two classes.
The canonical states are of the form of Eq.({\ref{1})  with $D(P)> 0$ and the extra state
\beq
P=\frac{1}{2}(e^1\wedge e^2\wedge e^3-e^1\wedge
e^{\overline{2}}\wedge e^{\overline{3}}-e^2\wedge
e^{\overline{3}}\wedge e^{\overline{1}}-e^3\wedge
e^{\overline{1}}\wedge e^{\overline{2}})
\label{Dnegativ}
\eeq
\noindent
with  $D(P)<0$.

\subsection{Embedded systems}

From the SLOCC classification of fermionic entanglement one can derive
other entanglement classes by restricting to a subgroup of the SLOCC group.
One way of achieving this is to constrain the set of admissible transformations
to ones that are also leaving invariant some extra structure.

In the special case of three fermions with six single particle states we can consider a fixed symplectic form $\omega\in \wedge^2V^{\ast}$ on $V={\mathbb C}^6$
and constrain the local operations to the set leaving $\omega$ invariant.
In this way we obtain the subgroup $GL(1,{\mathbb C})\times SP(6,{\mathbb C})
\subset GL(6,{\mathbb C})$. If we restrict 
the SLOCC group $GL(6,{\mathbb C})$
to this group the $20$ dimensional representation space decomposes to the direct sum of a $14$ and a $6$ dimensional
representation irreducible under $Sp(6,{\mathbb C})$.
\beq
\wedge^3 V^{\ast}=\omega\wedge V^{\ast}\oplus \wedge_0^3V^{\ast}.
\label{szimplconstr}
\eeq
\noindent
Here $\wedge_0^3V^{\ast}$ refers to the space of primitive three forms $P$ satisfying $\omega\wedge P=0$.
Chosing the fixed symplectic form as the one
\beq
\omega=e^1\wedge e^4+e^2\wedge e^5+e^3\wedge e^6=e^1\wedge e^{\overline{1}}
+e^2\wedge e^{\overline{2}}+e^3\wedge e^{\overline{3}}
\eeq
\noindent
one can see that the constraint $\omega\wedge P=0$
yields the one
\beq
P_{a14}+P_{a25}+P_{a36}=0,\qquad 1\leq a\leq 6.
\label{constraint11}
\eeq
\noindent
In the language of the $3\times 3$ matrices of Eq.(\ref{ymatr})-(\ref{Xmatr}) this means
that
\beq
X^t=X,\qquad Y=Y^t.
\eeq
\noindent
Taken together with $\eta$ and $\xi$ of Eq.(\ref{etaxi}) we obtain
the 1+6+6+1=14 independent components
of a three-form in
$\wedge_0^3V^{\ast}$.

The entanglement classes under the restricted group of admissible transformations turn out to be of the same structure then the ones under $GL(6,{\mathbb C})$. The orbit corresponding to the GHZ class is again a stable orbit.
This property dates back to the fact that the system we have considered is a PV which is class No.14. in the Sato-Kimura classification\cite{Satokimura}.
The classification of real entanglement classes is more involved\cite{Banos}.
For an explicit list see the appendix of the paper of Bryant in the first of Ref.\cite{Bryant}.

One can even restrict further the SLOCC group $GL(6,{\mathbb C})$
by regarding $V$ as the direct sum of three two dimensional
complex vector spaces. In this case we
have $V={\mathbb C}^6=V_1\oplus V_2\oplus V_3$. Let us furnish
$V_j, j=1,2,3$ with the symplectic forms $\omega_j\equiv e^j\wedge
e^{\overline{j}}$ and demand that the admissible set of
transformations are the ones leaving the symplectic forms {\it one
by one} invariant. This means that there is a group action
$Sp(2,{\mathbb C})^{\oplus 3}\simeq SL(2,{\mathbb C})^{\oplus 3}$ on
$V=V_1\oplus V_2\oplus V_3$. Taken together with an overall
complex rescaling we obtain the SLOCC group
$GL(2,{\mathbb C})^{\oplus 3}$. Now in this way we obtain the
constraints $P_{a14}=P_{a25}=P_{a36}=0, 1\leq a\leq 6$. This means
that only the $8$ amplitudes
$P_{123},P_{12{\overline{3}}},P_{1\overline{2}3},\dots
P_{\overline{123}}$ are surviving.
In this way the labels of the $V_j$ can be
mapped to the labels of three {\it distinguishable} qubits.
Labelling the qubits from the left to the right under the
correspondence \beq
(P_{123},P_{12{\overline{3}}},P_{1\overline{2}3},\dots
P_{\overline{123}})\leftrightarrow
(\psi_{000},\psi_{001},\psi_{010},\dots, \psi_{111})
\label{embedding}\eeq \noindent and a similar one for the basis
vectors ($e^1\wedge e^2\wedge e^3\mapsto \vert 000\rangle$ etc.)
our special three-form can be mapped to a three qubit state with
the usual SLOCC group $GL(2,{\mathbb C})^{\times 3}$ acting on it. Now the labels $1,2,3$ are referring
to the labels of the distinguishable constituents, on the other
hand numbers without an overline correspond to "$0$" and ones with
a overline correspond to "$1$". Notice also that in this case the
invariant ${\cal D(P)}$ of Eq.(\ref{Cayleygen}) restricts to
Cayley's hyperdeterminant $D(\psi)$ as given by Eq.(\ref{Cayley}).

This example can be generalized for different possible splits of
$V={\mathbb C}^6$ with the result of different special entangled
systems\cite{levvran2}. All of them and their corresponding
restricted sets of SLOCC transformations are embedded into $V$ and
the basic $GL(6,{\mathbb C})$ action on it. One particular example
that we need later can be obtained as follows. Let us consider the
embedding as given by Eq.(\ref{embedding}) and as a further
restriction demand that the admissible transformation are
consisting of the action of the {\it same} $S\in GL(2,{\mathbb
C})$ on each of the $V_j$s. In the three-qubit picture this means
that the restricted SLOCC group now acts as \beq
\vert\psi\rangle\mapsto (S\otimes S\otimes
S)\vert\psi\rangle,\qquad S\in GL(2,{\mathbb C}) \eeq \noindent
with $\vert\psi\rangle$ of the form \beq \vert\psi\rangle=
\psi_{000}\vert 000\rangle+\dots +\psi_{111}\vert
111\rangle,\qquad \psi_{001}=\psi_{010}=\psi_{110},\quad
\psi_{110}=\psi_{101}=\psi_{011}.
 \eeq \noindent
Hence in this case the number of independent complex amplitudes is
$4$ and the representation space for the $GL(2,{\mathbb C})$
action is the symmetrized tensor product of three ${\mathbb
C}^2$s. Clearly this situation is describing three
indistinguishable {\it bosonic} qubits. The relative invariant
which is the entanglement measure characterizing this situation is
a convenient truncation of Cayley's hyperdeterminant of
Eq.(\ref{Cayley}) \beq
d(x)=x_1^2x_4^2-6x_1x_2x_3x_4+4x_1x_3^3+4x_2^3x_4-3x_1^2x_4^2
\eeq \noindent where \beq x_1=\psi_{000},\quad
x_2=\psi_{001},\qquad x_3=\psi_{110},\qquad x_4=\psi_{111}. \eeq
\noindent
 This example gives rise to a particular PV
called No.4 in the Sato-Kimura classification scheme of regular
PVs \cite{Satokimura}.

Notice however, that all of our embedded systems were based on the $V={\mathbb C}^6$ case which is very special.
It is based on the Freudenthal system related to the cubic Jordan algebra of
$3\times 3$ complex matrices\cite{Krutelevich}.
This can be regarded as the "complexification"\cite{Krutelevich} of
he Jordan algebra of $3\times 3$ Hermitian matrices.
According to Eqs.(\ref{etaxi})-(\ref{ymatr}) this gives $1+9+9+1=20$ components for the corresponding Freudenthal triple
system.
This system also gives rise to a PV.
On the other hand the prehomogeneous vector space we encountered in the beginning of this subsection is related to the complexification of the simple Euclidean Jordan algebra of rank three based on the $3\times 3 $ symmetric matrices.  This gives rise to $1+6+6+1=14$ components for the corresponding Freudenthal triple system.
There are two more PVs of that type.
These are the No. 23 and 29 classes in the Sato-Kimura classification\cite{Satokimura}.
They are related to Freudenthal systems based on the compexifications of the Jordan algebras
of $3\times 3$ quaternion and octonian Hermitian matrices\cite{Krutelevich}.
They give rise to PVs with the correponding splits and dimensions:
$1+15+15+1=32$ and $1+27+27+1=56$.
One expects that the stable orbits of these PVs should give rise to GHZ-like classes and some particular relative invariants that can be used as entanglement measures.
Before justifying our expectations we have to turn our attention to string theory, the field where these exotic structures have first been applied.

\section{Hitchins functional}
\subsection{Hitchin's invariant as an entanglement measure}

Let us consider the {\it real vector space} $W={\mathbb R}^6$ and
the three-form $\varrho\in \wedge^3W^{\ast}$. Then after
introducing the $6\times 6$ matrix \beq
{(K_{\varrho})^a}_b=\frac{1}{2!3!}\varepsilon^{ai_2i_3i_4i_5i_6}{\varrho}_{bi_2i_3}{\varrho}_{i_4i_5i_6}
\label{KP} \eeq \noindent Hitchin's invariant\cite{Hitchin} can be
expressed as \beq
\lambda(\varrho)=\frac{1}{6}{(K_{\varrho})^a}_b{(K_{\varrho})^b}_a.
\label{Hitchinform} \eeq \noindent Clearly after identifying a
{\it real} three-form $P\in\wedge^3W^{\ast}$ from the previous
subsection with $\varrho$ and using Eq.(\ref{Cayleygen}) one
obtains \beq {\lambda}(\varrho)={\cal D}(\varrho).
\label{ekviv}\eeq \noindent Hence Hitchin's invariant
$\lambda(\varrho)$ is just our relative invariant used as an
entanglement measure in the previous section.

Regarding as an endomorphism of $V$ one can write
$K_{\varrho}={(K_{\varrho})^a}_be^b\otimes e_a$. It is
known\cite{Hitchin} that ${\rm Tr}K_{\varrho}=0$, hence
$K_{\varrho}\in sl(6,{\mathbb R})\subset gl(6,{\mathbb R})$. As a
Lie-algebra element with trace zero $(K_{\varrho})$ acts on a
three form as\cite{Gualtieri} (see also the Appendix in this
respect) \beq K_{\varrho}\cdot\varphi=-\frac{1}{2}{\rm
Tr}(K_{\varrho})\varphi+ {({K_{\varrho}})^a}_be^b\wedge
i_{e_a}\varphi=(K_{\varrho})^{\ast}\varphi . \eeq \noindent then
the correpondence between the two alternative ways of describing
${\cal D}(\varrho)$ namely the one of Eq.(\ref{quarticinvper2})
and Eq.(\ref{Hitchinform}) is given by \beq
\tilde{\varrho}=\frac{1}{3}(K_{\varrho})^{\ast}\varrho \eeq
\noindent giving rise to the formula
$\tilde{\varrho}_{abc}={\varrho}_{dbc}{(K_{\varrho})^d}_a$.

Notice also that for real three-forms $\varrho$ by virtue of
Eq.(\ref{quarticinvper2}) after employing the Freudenthal dual of
Eq.(\ref{freudual}) one can write \beq 2{\rm sgn}({\cal
D})\sqrt{\vert{\cal
D}(\varrho)\vert}\epsilon=\varrho\wedge\hat{\varrho}(\varrho)
\label{Hamiltoni1} \eeq \noindent where $\epsilon=e^1\wedge
e^2\wedge e^3\wedge e^4\wedge e^5\wedge e^6$. We recall that \beq
{\cal D}(\hat{\varrho})={\cal D}(\varrho) \eeq \noindent and
 \beq
 \hat{\hat{\varrho}}=-\varrho. \label{antiinv} \eeq \noindent It is
 important to realize that the latter three identities are
 satisfied for {\it all} Freudenthal triple systems\cite{freuddual}
 not merely the ones related to three-forms.

Our result is that Hitchin's nonlinear function
$\hat{\varrho}(\varrho)$ is the Freudenthal dual of $\varrho$.
Apart from giving an explicit formula for $\hat{\varrho}$ this result also
elucidates many of the important formulae obtained in
Ref.\cite{Hitchin}. Notice for example that Eq.(11) of that paper
is just a special case of our Eq.(\ref{Hamiltoni1}) when
$\lambda({\varrho})={\cal D}(\varrho)<0$. Moreover, for {\it both}
of the two real GHZ-like entanglement classes the important
identity Eq.(\ref{antiinv}) holds. For these two classes the forms
\beq \alpha=\varrho+\hat{\varrho}(\varrho),\quad
\beta=\varrho-\hat{\varrho}(\varrho),\qquad {\cal D}(\varrho)>0
\eeq \noindent and \beq \Omega=\varrho +
i\hat{\varrho}(\varrho),\quad \overline{\Omega}=\varrho -
i\hat{\varrho}(\varrho),\qquad {\cal D}(\varrho)<0 \label{kanform1}:\eeq \noindent
are belonging to the fully separable entanglement
class\cite{Hitchin}. In either case we have a two term
decomposition for $\varrho$ namely $\varrho=(\alpha+\beta)/2$ and
$\varrho =(\Omega+\overline{\Omega})/2$ which is up to
normalization of the canonical GHZ form. (In the case of the
three-qubit embedding just have a look at Eqs.(\ref{GHZ}) and
(\ref{ketto}).) This trick clearly also works in the complex case
hence we have an explicit metod for calculating the canonical form
of entangled states belonging to the stable orbit. For ${\cal D}<0$ via the
property\cite{Hitchin} $K_{\varrho}^2={\cal D}(\varrho)1$ the
$6\times 6$ matrix $I_{\varrho}=K_{\varrho}/\sqrt{-{\cal
D}(\varrho)}$ defines a complex structure on $W$. With respect to
this complex structure $\Omega$ is of type $(3,0)$.

Another important property that the Freudenthal formalism
automatically takes care is the nice symplectic geometry on the
space of real three forms\cite{Hitchin}. The phase space is
$\wedge^3W^{\ast}$ with the symplectic form as defined by
Eq.(\ref{sympl}). According to Eq.(\ref{Hamiltoni1}) one can see
that $H(\varrho)=\sqrt{\vert{\cal D}(\varrho)\vert}$ can be
regarded as the Hamiltonian and the Hamiltonian vector field
$X_H=\mp \hat{\varrho}(\varrho)$ for ${\rm sgn}({\cal D})=\mp 1$
i.e. up to sign it is just the Freudenthal dual of $\varrho$.
Moreover one can see that $K_{\varrho}$ is related to the moment
map. Moreover, for the special case of real three-forms with
${\cal D}(\varrho)<0$ according to Proposition 5. of
Ref.\cite{Hitchin} the derivative of $-\hat{\varrho}$ at $\varrho$
defines an integrable complex structure $J_{\varrho}$ on the
corresponding open orbit of stable forms. Note that we already have
a complex structure on $W$ defined by $I_{\varrho}$ according to
which we have the decomposition \beq \wedge^3W^{\ast}\otimes
{\mathbb C}=\wedge^{3,0}\oplus \wedge^{2,1}\oplus
\wedge^{1,2}\oplus \wedge^{0,3} \eeq \noindent One can clarify the
relationship between $J_{\varrho}$ and $I_{\varrho}$ by checking
the action of $J_{\varrho}$ on the type decomposition above. The
result is that $J_{\varrho}$ acts as $i$ on $\wedge^{3,0}\oplus
\wedge^{2,1}$ and  as $-i$ on $\wedge^{1,2}\oplus \wedge^{0,3}$.

Generally since the symplectic properties rest on the ones of
Freudenthal triple systems we can regard these as nice examples of
classical mechanical systems. Then the symplectic form is the
usual one defined for such systems and the square root of the
magnitude of the quartic invariant\cite{Krutelevich} is playing
the role of the Hamiltonian. The Freudenthal dual in all cases can
then be regarded as the Hamiltonian vector field. This observation
will be playing some role later.

\subsection{Hitchin's functional and semiclassical black hole entropy}

Let us now consider a closed oriented $6$-manifold $M$ and a real
three-form $\varrho$ with local coordinates $x^a$ in a coordinate
patch expressed as \beq
\varrho=\frac{1}{3!}{\varrho}_{abc}(x)dx^a\wedge dx^b\wedge
dx^c\in\wedge^3T^{\ast}M. \eeq \noindent Then Hitchin's functional
is defined as \beq V_H(\varrho)=\int_{M}\sqrt{\vert{\cal
D}(\varrho)\vert}d^6x\label{Hitchinfunk} \eeq \noindent where
${\cal D}(\varrho)$ related to our entanglement measure of
Eq.(\ref{tanglegen}) is defined by either Eq.(\ref{Cayleygen})
with $P$ replaced by $\varrho$ or by
Eqs.(\ref{Hitchinform})-(\ref{ekviv}). Using the observation that
$\hat{\varrho}$ is the Freudenthal dual of $\varrho$ by virtue of
Eq.(\ref{Hamiltoni1}) an alternative formula for this functional
is \beq V_H(\varrho)=\frac{1}{2}{\rm sgn}({\cal
D(\varrho)})\int_{M}\varrho\wedge\hat{\varrho}(\varrho).\label{Hitchinfunk2}
\eeq \noindent

In the special case when ${\cal D(\varrho)}<0$ everywhere on $M$,
each differential three-form $\varrho$ defines an almost complex
structure $I_{\varrho}$ on $M$. If $\varrho$ is a critical point
of $V_H(\varrho)$ on a cohomology class of $H^3(M,{\mathbb R})$
($d\varrho =0$) then it follows\cite{Hitchin} that we also have
$d\hat{\varrho}=0$. Hence the separable three-form
$\Omega=\varrho+i\hat{\varrho}(\varrho)$ of type $(3,0)$
introduced in the previous subsection is closed and the almost
complex structure $I_{\varrho}$ defined by $\varrho$ is
integrable. As a result of these considerations a critical point
or a classical solution of $V_H(\varrho)$ defines a complex
structure on $M$ with a non-vanishing holomorphic three-form
$\Omega$. Note that in terms of
$\Omega=\varrho+i\hat{\varrho}(\varrho)$ Hitchin's functional is
just the holomorphic volume of $M$ \beq
V_H(\varrho)=-\frac{i}{4}\int_M\Omega\wedge\overline{\Omega}.
\label{holomorphic}\eeq\noindent

 In particular Calabi-Yau three-folds used by string
theorists in models of string compactification are K\"ahler
manifolds with a nonvanishing holomorphic three-form $\Omega$.
Hence the {\it complex structure} of such manifolds can be derived
from the critical points of $V_H(\varrho)$. The phenomenon of
obtaining a particular complex structure from a fixed three-form
$\varrho$ also occurs in the case of $4D$ BPS black holes in Type
IIB string theory compactified on  Calabi-Yau three-folds via the
attractor mechanism\cite{Becker,FKS}. In this case fixing the BPS
charge configuration of the black hole solution amounts to fixing
a homology class $\gamma\in H_3(M,{\mathbb Z})$ corresponding to a
wrapping configuration of $3$-branes. The cohomology class of
$\varrho$ then equals the Poincar\'e dual $\Gamma\in
H^3(M,{\mathbb Z})$ of $\gamma$. The attractor mechanism provides
a particular holomorphic three-form $\Omega$ at the black hole
horizon in terms of the charges. Identifying the {\it real part}
of  $\Omega$ with $\varrho$ the attractor mechanism gives the
imaginary part $\hat{\varrho}$ in terms of $[\varrho]=\Gamma$.

This argument has been suggested in Ref.\cite{Dijkgraaf} to relate
the value of $V_H(\varrho)$ at the critical point to the
semiclassical Bekenstein-Hawking entropy. Since the main
correspondence of the BHQC is the one existing between the
semiclassical black hole entropy and certain entanglement measures
it is instructive to revisit this argument in the context of the
BHQC using type IIB string theory. In type IIB compactification on
a CY three-fold $M$ with holomorphic three-form $\Omega$ the
resulting low energy theory is four-dimensional $N=2$
supergravity. In this theory we have $h^{2,1}(M)$ vector
multiplets. Let us denote by $X^I, I=1,2,\dots h^{2,1}$ the scalar
components of these multiplets describing the {\it complex
structure} moduli of $M$. The vector multiplet part of the
effective action is fully specified by the holomorphic
prepotential ${\cal F}(X)$ defining a special K\"ahler geometry
(with K\"ahler potential $K$) of the moduli space of $M$. Denoting
$F_I={\partial}_I{\cal F}$ we have \beq
X^I=\int_{A^I}\Omega,\qquad F_I=\int_{B^I}\Omega \eeq \noindent
\beq \Omega=X^I\alpha_I-F_I(X)\beta^I \eeq \noindent and \beq
K=-\log i(\overline{X}^IF_I-X^I\overline{F}_I). \label{Kpot} \eeq
\noindent
 Here
$\{A^I,B_I\}$ form a basis for the three-cycles in $H_3(M,{\mathbb
Z})$ and $\{\alpha_I,\beta^I\}$ are the dual basis three-forms of
$H^3(M,{\mathbb Z})$. In this setting the holomorphic volume is
\beq V_H(\varrho)=\frac{1}{4i}\int_M\Omega\wedge
\overline{\Omega}=\frac{1}{2}{\rm
Im}(X^I\overline{F}_I)=\frac{1}{4}e^{-K}.
\label{kellenifog}\eeq\noindent Let us introduce for a $\gamma \in
H_3(M,{\mathbb Z})$ its Poincar\'e dual $\Gamma$ as \beq
\Gamma=p^I\alpha_I-q_I\beta^I. \eeq \noindent Then the central
charge field is \beq Z(\gamma)=e^{K/2}\int_{\gamma}\Omega
=e^{K/2}\int_M\Omega\wedge\Gamma=e^{K/2}(p^IF_I-q_IX^I).\eeq
\noindent One can show\cite{Becker} that for static spherically
symmetric extremal BPS black hole solutions the semiclassical
black hole entropy is \beq S=\pi\vert Z\vert^2=\pi\frac{\vert
p^IF_I-q_IX^I\vert^2}{2{\rm Im}X^I\overline{F}_I}.
\label{entropia}\eeq \noindent Here it is understood that $S$ is
depending on the moduli $X^I$ and the charges $p^I,q_I$, moreover
the values of the moduli fields should be taken at the black hole
horizon. According to the attractor mechanism\cite{FKS,Becker}
these values for the moduli can be expressed in terms of the
charges via the attractor equations \beq {\rm Re}(CX^I)=p^I,\qquad
{\rm Re}(CF_I)=q_I,\qquad C=-2i\overline{Z}e^{K/2}.
\label{attractoreq} \eeq \noindent Since the formula for the
entropy is invariant under dilatations one can set $C=1$ in
Eq.(\ref{attractoreq}). As a consequence of this the charges are
just the {\it real parts} of the quantities of $X^I$ and $F_I$
hence after putting this into Eq.(\ref{entropia}) we get \beq
S=\frac{\pi}{2}{\rm Im}(X^I\overline{F}_I).
\label{ent}\eeq\noindent Here the charges via
Eq.(\ref{attractoreq}) also determine the {\it imaginary} parts of
$X^I$ and $F_I$, hence $S$ can be expressed entirely in terms of
the charges $p^I$ and $q_I$.

Let us now compare this implicit expression for the entropy as
given by Eq.(\ref{ent}) and using Eq.(\ref{kellenifog}) the
similar expression for Hitchin's functional of
Eq.(\ref{holomorphic}). At the critical point of $V_H(\varrho)$
where $\varrho$ determines the imaginary part $\hat{\varrho}$ of
$\Omega$ we clearly have \beq S_{BH}=\pi V_H(\varrho_{\rm
crit}),\qquad [\varrho]=\Gamma.\label{lenyeg}\eeq \noindent This establishes
the a link between the value of the extremized action
$V_{H}(\varrho)$ based on an entanglement measure ${\cal
D}(\varrho)$ and the semiclassical black hole entropy.

\subsection{An example: $T^6$  }

In order to elucidate the meaning of Eq.(\ref{lenyeg}) we consider
as our oriented closed $6$-manifold the torus $T^6$. We choose
real coordinates $u^i, v^i, i=1,2,3$ and the orientation
$\int_{T^6} du^1\wedge dv^1\wedge du^2\wedge dv^2\wedge du^3\wedge
dv^3 =1$. Next we consider a wrapping configuration $\gamma\in
H_3(T^6,{\mathbb Z})$ and we expand its Poincar\'e dual $\Gamma\in
H^3(T^6,{\mathbb Z})$ in the basis satisfying
$\int_{T^6}\alpha^I\wedge \beta_J=\delta^I_J,\quad I,J=1,2,\dots
10$ \beq \alpha_0=du^1\wedge du^2\wedge du^3,\qquad
\alpha_{ij}=\frac{1}{2}{\varepsilon}_{ii^{\prime}j^{\prime}}du^{i^{\prime}}\wedge
du^{j^{\prime}}\wedge dv^j \eeq \noindent \beq \beta^0=-dv^1\wedge
dv^2\wedge dv^3,\qquad
\beta^{ij}=\frac{1}{2}{\varepsilon}_{ji^{\prime}j^{\prime}}du^i\wedge
dv^{i^{\prime}}\wedge dv^{j^{\prime}} \eeq\noindent as \beq
\Gamma=p^0\alpha_0+P^{ij}\alpha_{ij}-Q_{ij}\beta^{ij}-q_0\beta^0.\eeq\noindent

We write the nondegenerate real three-form $\varrho$ belonging to the class with ${\cal D}(\varrho)<0$ featuring Hitchin's functional $V_H(\varrho)$ as \beq \varrho=\sum_{1\leq a<b<c\leq
6}\varrho_{abc}f^a\wedge f^b\wedge f^c
\label{fermistate}\eeq\noindent where \beq
(f^1,f^2,f^3,f^4,f^5,f^6)\equiv
(f^1,f^2,f^3,f^{\overline{1}},f^{\overline{2}},f^{\overline{3}})=(du^1,du^2,du^3
,dv^1,dv^2,dv^3).\eeq\noindent
According to Eq.(\ref{lenyeg}) up to an exact form we should identify $\varrho$  with
$\Gamma$.
Explicitly this identification is given by the
expressions \beq p^0=\varrho_{123},\qquad
\begin{pmatrix}P^{11}&P^{12}&P^{13}\\
               P^{21}&P^{22}&P^{23}\\
                      P^{31}&P^{32}&P^{33}\end{pmatrix}
              =\begin{pmatrix}\varrho_{23\overline{1}}&\varrho_{23\overline{2}}&\varrho_{23\overline{3}}\\
              \varrho_{31\overline{1}}&\varrho_{31\overline{2}}&\varrho_{31\overline{3}}\\
              \varrho_{12\overline{1}}&\varrho_{12\overline{2}}&\varrho_{12\overline{3}}\end{pmatrix},
\label{pma}\eeq\noindent

\beq q^0=\varrho_{\overline{1}\overline{2}\overline{3}},\qquad
\begin{pmatrix}Q_{11}&Q_{12}&Q_{13}\\
               Q_{21}&Q_{22}&Q_{23}\\
                          Q_{31}&Q_{32}&Q_{33}\end{pmatrix}
                  =\begin{pmatrix} \varrho_{1\overline{2}\overline{3}}&
                  \varrho_{1\overline{3}\overline{1}}&
                  \varrho_{1\overline{1}\overline{2}}\\
                  \varrho_{2\overline{2}\overline{3}}&
                  \varrho_{2\overline{3}\overline{1}}&
                  \varrho_{2\overline{1}\overline{2}}\\
                  \varrho_{3\overline{2}\overline{3}}&
                  \varrho_{3\overline{3}\overline{1}}&
                  \varrho_{3\overline{1}\overline{2}}\end{pmatrix}.\label{qma}\eeq\noindent

Now a critical point of $V_H(\varrho)$ gives rise to a fully
separable state of the form
$\Omega=\varrho+i\hat{\varrho}(\varrho)$ where $\hat{\varrho}$ is
the Freudenthal dual of $\varrho$ expressed in terms of the
charges. For $\hat{\varrho}$ one can use Eq.(\ref{freudual}) or
the formulae \beq \hat{p}^0=\frac{-\tilde{p}^0}{\sqrt{-{\cal
D}}},\quad \hat{P}=\frac{-\tilde{P}}{\sqrt{-{\cal D}}} \eeq
\noindent \beq \hat{q}^0=\frac{-\tilde{q}^0}{\sqrt{-{\cal
D}}},\quad \hat{Q}=\frac{-\tilde{Q}}{\sqrt{-{\cal D}}} \eeq
\noindent valid for {\it all} Freudenthal triple systems. Here
\beq {\cal D}=[p^0q_0-(P,Q)]^2-
 4(P^{\sharp},Q^{\sharp})+4p^0
  N(Q)+4q_0N(P), \label{Cayleygen22}
   \eeq
\beq
\tilde{p}^0=-2N(P)-p^0(p^0q_0-(P,Q)),
\qquad
\tilde{P}=2(p^0Q^{\sharp}-Q\times P^{\sharp})
-(p^0q_0-(P,Q))P
\label{tild1}
\eeq
\noindent
\beq
\tilde{q}^0=2N(Q)+q^0(p^0q_0-(P,Q)),
\qquad
\tilde{Q}=-2(q^0P^{\sharp}-P\times Q^{\sharp})
+(p^0q_0-(P,Q))Q.
\label{tild2}
\eeq
\noindent
Here in our special case $(A,B)={\rm Tr}(AB)$ and $N(A)={\rm Det}(A)$
for the remaining definitions see the Appendix.

Now this particular $\Omega$ arising from the
critical point of $V_{H}(\varrho)$ can be expanded as
\beq
 \Omega=C\Omega_0=C\left(\alpha_0+{\tau}^{jk}\alpha_{jk}+{\tau^{\sharp}}_{jk}\beta^{kj}-({\rm
 Det}\tau)\beta^0\right),\label{kifejtes2}\eeq\noindent
Where we put back the factor $C$ of Eq.(\ref{attractoreq}).  One
can then introduce complex coordinates \beq z^i=u^i+\tau^{ij}v^j
\eeq \noindent such that the separable form is manifest \beq
\Omega=C\Omega_0=C dz^1\wedge dz^2\wedge
dz^3=\varrho+i\hat{\varrho}(\varrho) \label{kifejtes3} \eeq
\noindent i.e. $\Omega_0$ is a holomorphic three-form for the
torus. Here for the expansion coefficients $\tau^{ij}$ fixing the
complex structure of $T^6$ we chose the convention \beq
\tau^{ij}=x^{ij}-iy^{ij},\qquad y^{ij}>0,\label{taumatrix} \eeq
\noindent One can also check
 that by virtue of Eq.(\ref{kellenifog})
\beq e^{-K}=8{\rm Det}y.\label{matrixkahler}\eeq\noindent

From Eqs.(\ref{kifejtes2}) and (\ref{kifejtes3}) one can see that
the complex structure obtained from the extremization of Hitchin's
functional is \beq \tau=\frac{P+i\hat{P}}{p^0+i\hat{p^0}}
\label{Kallosh2} \eeq \noindent or after performing standard
manipulations\cite{Moore,qfromextra} using identitities for
Freudenthal systems \beq
\tau=\frac{1}{2}\left[-(2PQ+[p^0q_0-(P,Q)])+i\sqrt{-{\cal
D}}\right](P^{\sharp}-p^0Q)^{-1}. \label{tauexpl} \eeq \noindent
According to Eq.(\ref{kellenifog}) the value of $V_H$ at the
critical point is $\frac{1}{4}\vert C\vert^2e^{-K}$ 
. Using this we obtain the final result \beq S_{BH}=\pi
V_H(\varrho_{crit})=\pi\sqrt{-{\cal D}} \label{entrasameasure}
\eeq \noindent where ${\cal D}$ is given by
Eq.(\ref{Cayleygen22}). This result shows that the semiclassical
black hole entropy is given by the entanglement measure ${\cal D}$
for the three-fermion state as given by
Eqs.(\ref{fermistate})-(\ref{qma}).

It is instructive to express $[\varrho]=\Gamma$ from
Eq.(\ref{kifejtes3}) in the form \beq \Gamma=
\frac{1}{2}(C\Omega_0+\overline{C}\overline{\Omega}_0)
=\overline{Z}(-ie^{K/2}\Omega_0)+(-Z)(-ie^{K/2}\overline{\Omega}_0).
\label{Hodge1} \eeq \noindent Let us introduce the Hermitian inner
product for three-forms as
\beq\langle\varphi\vert\psi\rangle=\int_{T^6}\varphi\wedge\ast\overline{\psi}
\label{inner} \eeq \noindent where $\ast$ is the Hodge star. One
can then regard $H^3(T^6,{\mathbb{C}})$ equipped with
$\langle\cdot\vert\cdot\rangle$ as a $20$ dimensional Hilbert
space. One can then see\cite{qfromextra} that Eq.(\ref{Hodge1})
can be written in the form \beq \vert
\Gamma\rangle={\Gamma}_{123}\vert 123\rangle
+{\Gamma}_{\overline{123}}\vert \overline{123}\rangle \eeq
\noindent where $\vert 123\rangle$ and
$\vert\overline{123}\rangle$ are orthonormal basis vectors. Since
$\vert Z\vert^2=\sqrt{-{\cal D}}$ one can show that \beq
\vert\Gamma\rangle=(-{\cal D})^{1/4}\left(e^{i\alpha}\vert
123\rangle -e^{-i\alpha}\vert\overline{123}\rangle\right),\qquad
\tan\alpha=\frac{p^0}{\hat{p}^0}. \label{GHZattr} \eeq \noindent
Notice that this "state" is of the GHZ-form, with the phase
factors are coming from the phase of the central charge expressed
in terms of the charges $p^0$ and its Freudenthal dual
$\hat{p}^0$. The quantity\cite{Levay3,qfromextra}
$\frac{1}{2}\pi\vert\vert\Gamma\vert\vert^2$ is just the
semiclassical entropy $S_{BH}$.

Now we employ an extra constraint
and consider our torus equipped with a
symplectic form $\omega$ giving rise to the volume form
compatible with the orientation,
and we also restrict $\varrho$ featuring Hitchin's functional
by the constraint
$\omega\wedge\varrho=0$.
In this case we have to find the constrained critical points\cite{stable} of $V_H$.
In the language of entanglement 
the new $\varrho$ arising from of Eq.(\ref{fermistate})
can  now  be regarded as an embedded real three-fermion state,
with the extra constraint restricting the SLOCC group from
$GL(6,{\mathbb R})$ to $GL(1,{\mathbb R}))\times SP(6,{\mathbb R})$.
According to Eq.(\ref{constraint11}) these considerations
give the restriction on the charge configuration
\beq
P^t=P,\qquad Q^t=Q\label{szimmetrikussag}
\eeq
\noindent
yielding $14$ independent charges.
Moreover, due to our restrictions corresponding to Eq.(\ref{szimmetrikussag}) $T^6$ with the arising complex structure will be a principally polarized Abelian variety with $\tau^{ij}=\tau^{ji}$.
The example we obtain in this way is just the one discussed by Moore\cite{Moore,qfromextra} when studying BPS attractor varieties in $IIB$ string theory compactified on $T^6$.
Black hole entropy is again given by Eq.(\ref{entrasameasure})
with the corresponding fomula is the one depending on $14$ charges. This is the square root of the quartic invariant of the Freudenthal triple system based on the cubic Jordan algebra of real $3\times 3$ symmetric matrices.

\subsection{STU truncation}

Let us now chose an $M$ having the product form $M=M_1\times M_2\times M_3$
where $M_{1,2,3}$ are two dimensional tori $T^2$ with coordinates $u^i,v^i$.
Here $i=1,2,3$ labels the different tori.
Now using the notation $e^i=du^i$, $e^{\overline{i}}=dv^i$ an element of $H^3(M,{\mathbb R})$ can be written as
\beq
{\varrho}={\varrho}_{123}e^1\wedge e^2\wedge e^3+{\varrho}_{\overline{1}23}
e^1\wedge e^2\wedge e^{\overline{3}}+\dots +{\varrho}_{\overline{123}}
e^{\overline{1}}\wedge e^{\overline{2}}\wedge e^{\overline{3}}
\eeq
\noindent
i.e. ${\varrho}$ has merely $8$ nonzero amplitudes.
It is convenient to relabel them in a notation reminiscent of the amplitudes of three-qubits
\beq
(\varrho_{123},\varrho_{12\overline{3}},\varrho_{1\overline{2}3},\dots,\varrho_{\overline{123}})\leftrightarrow (\varrho_{000},\varrho_{001},\varrho_{010},\dots,\varrho_{111}).
\eeq
\noindent
In the language of embedded systems (see Section 2.3) one can obtain this case
from the results of the previous subsection by employing the constraint
$\omega_i\wedge\varrho=0$, where $\omega_i$ are the symplectic forms of the tori.

Now a calculation  shows that
$K_{\varrho}$ has the form

\beq
{(K_{\varrho})^a}_b=\begin{pmatrix}U_{\varrho}&0&0\\0&T_{\varrho}&0\\0&0&S_{\varrho}\end{pmatrix},\qquad  a,b=1,\overline{1},2,\overline{2},3,\overline{3}
\eeq
\noindent
where
\beq
S_{\varrho}=\begin{pmatrix}({\varrho}_0\cdot {\varrho}_1)_1&({\varrho}_1\cdot {\varrho}_1)_1\\-({\varrho}_0\cdot{\varrho}_0)_1&-(\varrho_0\cdot\varrho_1)_1\end{pmatrix},\quad
T_{\varrho}=\begin{pmatrix}({\varrho}_0\cdot {\varrho}_1)_2&({\varrho}_1\cdot {\varrho}_1)_2\\-({\varrho}_0\cdot{\varrho}_0)_2&-(\varrho_0\cdot\varrho_1)_2\end{pmatrix}
\eeq
\noindent
\beq
U_{\varrho}=\begin{pmatrix}({\varrho}_0\cdot {\varrho}_1)_3&({\varrho}_1\cdot {\varrho}_1)_3\\-({\varrho}_0\cdot{\varrho}_0)_3&-(\varrho_0\cdot\varrho_1)_3\end{pmatrix}
\eeq
\noindent
Here for $A,B\in {\mathbb R}^4$ an $SL(2,{\mathbb R})\times SL(2, {\mathbb R})$
invariant inner product is defined as
\beq
A\cdot B=A_1B_4-A_2B_3-A_3B_2+A_4B_1
\label{dotproduct}
\eeq
\noindent
and the subscripts $1,2,3$  of the innner products mean that the splitting of the $8$ amplitudes into two four component vectors is effected by assigning to
qubit $1,2,3$ a special role.
Hence for example for calculating $(\varrho_0\cdot\varrho_1)_1$ the four component vectors to be used  in Eq.(\ref{dotproduct}) are
\beq
{\varrho}_0=\begin{pmatrix}\varrho_{000}\\\varrho_{001}\\\varrho_{010}\\\varrho_{011}\end{pmatrix},\qquad
{\varrho}_1=\begin{pmatrix}\varrho_{100}\\\varrho_{101}\\\varrho_{110}\\\varrho_{111}\end{pmatrix}\qquad
\label{twistors}
\eeq
\noindent
where now the first qubit (labelling the qubits from the left to the right) plays a special role.

In this notation Hitchin's invariant is
\beq
D(\varrho)=(\varrho_0\cdot\varrho_1)_n^2-(\varrho_0\cdot\varrho_0)_n(\varrho_1\cdot\varrho_1)_n, \qquad \forall n=1,2,3
\label{trialitycayley}
\eeq
\noindent
which is just another form for Cayley's hyperdeterminant
of Eq.(\ref{Cayley}) related to the entanglement measure the three-tangle.
Notice that the independence of $D(\varrho)$ of the particular split expresses the permutation invariance of $D$.
After these considerations one can immediately check that\cite{Hitchin}
\beq
{\rm Tr}K_{\varrho}=0,\qquad K_{\varrho}^2=D(\varrho){\bf 1}
\eeq
\noindent
where ${\bf 1}$ is the $6\times 6$ identity matrix.

For ${\cal D}(\varrho)<0$ the GHZ components $\Omega$ and $\overline{\Omega}$ of such a $\varrho$ are given
by Eq.(\ref{kanform1})
where now the Freudenthal dual $\hat{\varrho}$ is
arising from the Freudenthal system based on the Jordan algebra of $3\times 3$ {\it diagonal} matrices.
There is a geometric description of three-qubit entanglement in terms of twistors\cite{Levay1}.
In this picture finding the canonical GHZ components of $\varrho$ amounts to finding the principal null directions (with respect to the symmetric bilinear form of Eq.(\ref{dotproduct})) of bivectors like
$\varrho_{0a}\varrho_{1b}-
\varrho_{0b}\varrho_{1a}$ formed from the four-vectors of Eq.(\ref{twistors}).
For a real bivector with ${\cal D}\neq 0$ we have two principal null directions.We have  either two real directions ${\cal D}>0$ or they are coming in complex conjugate pairs if
${\cal D}<0$.
These cases correspond to the two inequivalent GHZ SLOCC classes.

Now for a three-form representing the cohomology class of a
wrapped $D3$ brane configuration we take  \beq
\Gamma=p^I\alpha_I-q_I\beta^I\in H^3(T^6,{\mathbb
Z})\label{threebrane}\eeq\noindent with summation on $I=0,1,2,3$
and \beq \alpha_0=du^1\wedge du^2\wedge du^3,\qquad
\beta^0=-dv^1\wedge dv^2\wedge dv^3\label{nullasok}\eeq\noindent
\beq \alpha_1=dv^1\wedge du^2\wedge du^3,\qquad \beta^1=du^1\wedge
dv^2\wedge dv^3\label{egyesek}\eeq\noindent with the remaining
ones obtained via cyclic permutation.

Let us now pretend that we have an {\it arbitrary} complex
structure on $M=T^2\times T^2\times T^2$. We introduce the
coordinates \beq z^j=u^j+\tau^jv^j,\qquad \tau^j=x^j-iy^j\qquad
y^j>0,\qquad j=1,2,3 \label{koordinatak}\eeq\noindent and the
holomorphic three-form \beq \Omega_0=dz^1\wedge dz^2\wedge
dz^3\label{holthree}\eeq\noindent with $\tau^j$ labelling the
complex structure. It is well-known\cite{Denef} that we can
express $\Gamma$ in a basis where the Hodge star is diagonal as
\beq \Gamma=e^{K/2}\left(iZ(\Gamma)\overline{\Omega}_0
-ig^{j\overline{k}}D_jZ(\Gamma)
\overline{D}_{\overline{k}}\overline{\Omega}_0+{\rm c.c.}\right)=
e^{K/2}\left(iZ(\Gamma)\overline{\Omega}_0
-i\delta^{\hat{j}\hat{\overline{k}}}D_{\hat{j}}Z(\Gamma)
\overline{D}_{\hat{\overline{k}}}\overline{\Omega}_0+{\rm
c.c.}\right)\label{expand}\eeq\noindent Here
$Z(\Gamma)=e^{K/2}\int_{T^6}\Gamma\wedge\Omega_0$ is the central
charge, and the flat covariant derivative in our special case 
is defined as \beq
D_{\hat{\tau}}\Omega_0\equiv(\overline{\tau}-\tau)D_{\tau}\Omega_0\equiv
(\overline{\tau}-\tau)\left({\partial}_{\tau}+{\partial}_{\tau}K
\right) \Omega_0,\label{covder}\eeq\noindent
where for simplicity we have omitted the labels of $\tau$. For the
STU truncation the explicit form of $Z(\Gamma)$ is \beq
Z(\Gamma)=e^{K/2}W(\tau^3,\tau^2,\tau^1)\label{centszuper}\eeq\noindent
where \beq W(\tau^3,\tau^2,\tau^1)=q_0+q_1\tau^1
+q_2\tau^2+q_3\tau^3+p^1\tau^2\tau^3+p^2\tau^1\tau^3+p^3\tau^1\tau^2-p^0\tau^1\tau^2\tau^3.
\label{centralcharge}\eeq\noindent Let us now restrict the
Hermitian inner product of Eq.(\ref{inner}) to the $8$ dimensional
untwisted primitive part of $H^3(T^6,{\mathbb C})$. One can then
introduce a Hodge diagonal basis in this space as
follows\cite{qfromextra}.  \beq -ie^{K/2}\Omega_0\leftrightarrow
\vert 000\rangle,\quad -ie^{K/2}D_{\hat{1}}\Omega_0\leftrightarrow
\vert 001\rangle, \quad
-ie^{K/2}D_{\hat{2}}\Omega_0\leftrightarrow \vert 010\rangle,\quad
-ie^{K/2}D_{\hat{3}}\Omega_0\leftrightarrow \vert
100\rangle\nonumber\label{qubitcorrespondence}\eeq\noindent \beq
-ie^{K/2}\overline{\Omega}_0\leftrightarrow \vert 111\rangle,\quad
-ie^{K/2}\overline{D}_{\hat{\overline{1}}}\Omega_0\leftrightarrow
\vert 110\rangle, \quad
-ie^{K/2}\overline{D}_{\hat{\overline{2}}}\Omega_0\leftrightarrow
\vert 101\rangle,\quad
-ie^{K/2}\overline{D}_{\hat{\overline{3}}}\Omega_0\leftrightarrow
\vert
011\rangle.\nonumber\label{qubitcorrespondence2}\eeq\noindent Now
with these definitions our $8$ dimensional space is isomorphic to
$({\mathbb C}^2)^{\times 3}$ equipped with a Hermitian inner
product, i.e. it is the space of states for three qubits. We must
note however, two important peculiarities. First of all the state
$ \Gamma\leftrightarrow\vert\Gamma\rangle$ defined as
the qubit version of Eq.(\ref{expand}) is complex in appearance.
However, its explicit form \beq
\vert\Gamma\rangle=\Gamma_{000}\vert 000\rangle +\Gamma_{001}\vert
001\rangle +\dots +{\Gamma}_{110}\vert 110\rangle
+{\Gamma}_{111}\vert 111\rangle\label{qubitform2}\eeq\noindent
where \beq
{\Gamma}_{111}=-e^{K/2}W(\tau^3,\tau^2,\tau^1)=-\overline{\Gamma}_{000}\label{ampl1}\eeq\noindent
\beq
{\Gamma}_{001}=-e^{K/2}W(\overline{\tau}^3,\overline{\tau}^2,\tau^1)=-\overline{\Gamma}_{110}
\label{ampl2},\qquad {\rm etc.} \eeq\noindent  shows that it
satisfies an extra reality condition coming from
Eq.(\ref{threebrane}).
 Second for BPS states with $D<0$ the
amplitudes ${\Gamma}_{000}$ and ${\Gamma}_{111}$ are obviously
playing a special role since they are connected to the holomorphic
structure via the appearance of $\Omega_0$ in Eq.(\ref{expand}).

These considerations show that in the case of the STU truncation
by putting an {\it arbitrary} complex structure on $M=T^2\times
T^2\times T^2$ one can define a charge and moduli dependent
three-qubit state of the (\ref{qubitform2}) form. The critical
point of Hitchin's functional on the other hand defines a ${\it
special}$ complex structure on $M$. It is given by equation
(\ref{tauexpl}) with keeping only the diagonal entries of the
$3\times 3$ matrices showing up in this formula. In this case our
three qubit state has the special form \beq
\vert\Gamma_{crit}\rangle=(-{ D})^{1/4}\left(e^{i\alpha}\vert
000\rangle -e^{-i\alpha}\vert\overline{111}\rangle\right),\qquad
\tan\alpha=\frac{p^0}{\hat{p}^0}. \label{GHZattrstu} \eeq
\noindent where now $D$ is Cayley's hyperdeterminant related to
the three-tangle of Eq.(\ref{threetangle}) as the canonical
entanglement measure for three-qubit systems. 
The Freudenthal dual component $\hat{p}^0$ is also modified accordingly.
The semiclassical
black hole entropy is given by \beq S_{BH}=\pi
V_H(\varrho_{crit})=\pi\sqrt{- D} \label{entrasameasure1} \eeq
\noindent where $D$ is given by Eq.(\ref{Cayley}). It is important
to realize that again
$S_{BH}=\frac{1}{2}\pi\vert\vert\Gamma_{crit}\vert\vert^2$. On the
other hand the norm squared of the state $\vert\Gamma\rangle$
featuring and arbitrary complex structure is the Black Hole
Potential\cite{stu,Saraikin,Levay1} \beq
V_{BH}=\frac{1}{2}\vert\vert\Gamma\vert\vert^2. \label{BHpot} \eeq
\noindent

For BPS black holes we have $D<0$. In this case studying the
explicit form of the full radial flow from the asymptotically
Minkowski region to the event horizon one can follow  the
transition from a three-qubit state of the form
Eq.(\ref{qubitform2}) to the one of the Eq.(\ref{GHZattrstu})
form. In the literature of the BHQC this process is called a
distillation procedure of a GHZ state. This process is terminated
at the horizon where $\Gamma_{001}=\Gamma_{010}=\Gamma_{100}=0$.
This is just the usual process well-known in the string theory
literature for which $\Gamma$ has only $H^{3,0}$ and $H^{0,3}$
components. This observation is originally due to
Moore\cite{Moore}. We must stress that the formalism based on
$V_H$ is more general.

We would like to close this subsection with an important comment
on {\it non-BPS} black holes\cite{NBPS}. In the most general
setting it is tempting to relate the critical points in the ${\cal
D}>0$ branch of real states of the Hitchin functional of the form
as given by Eq.(\ref{Hitchinfunk2}) with the critical points of
the general expression for the Black Hole
Potential\cite{Saraikin}. It is easy to see that using the
Hermitian inner product of Eq.(\ref{inner}) the "norm squared"
interpretation of Eq.(\ref{BHpot}) survives even in this case. In
the special case of the STU truncation studied here the explicit
form of these solutions is known\cite{stu}. In the framework of
the BHQC the non-BPS analogues of the state of
Eq.(\ref{GHZattrstu}) are again of special form. They are
belonging to the GHZ-class with $D>0$ and called graph states in
the QI literature\cite{Levay3}. The interpretation for the
attractor mechanism as being some sort of distillation procedure
also works in this case.\cite{Levay3} An especially interesting
feature of these solutions is that for the non-BPS branch the
special role of the holomorphic three-form in the
expansion of Eq.(\ref{qubitform2}) is lost. This can be seen most
clearly for non-BPS solutions giving rise to attractors with
vanishing central charge\cite{vanishing}. In this case
${\Gamma}_{000}$ and ${\Gamma}_{111}$ i.e. precisely the canonical
GHZ amplitudes are vanishing. Hence also including the non-BPS
branch into the picture renders the interpretation of $\Gamma$ as
a quantity relating to some sort of "state" more natural. We will
have something more to say on these interesting issues in Section
6.

\subsection{$t^3$ truncation}

The $t^3$ truncation is the diagonal torus example where
$M=T^2\times T^2\times T^2$ with the tori regarded
indistinguishable. In the
entanglement picture this case amounts to considering three
indistinguishable bosonic qubits. Now $\Gamma\in H^3(M,{\mathbb
Z})$ is expanded as \beq \Gamma=
p^0\alpha_0+p(\alpha_1+\alpha_2+\alpha_3)
-q(\beta^1+\beta^2+\beta^3)-q_0\beta^0. \eeq \noindent In the
Hodge diagonal basis we have
${\Gamma}_{001}={\Gamma}_{010}={\Gamma}_{100}$ etc. hence our
charge and moduli dependent state will be of the form \beq
\vert\Gamma\rangle=\Gamma_{000}\vert 000\rangle+\Gamma_{001}\left
(\vert 001\rangle+\vert 010\rangle+\vert 100\rangle\right)+
\Gamma_{110}\left(\vert 110\rangle+\vert 101\rangle+\vert
011\rangle\right)+\Gamma_{111}\vert 111\rangle. \eeq \noindent The
critical point of Hitchin's functional gives the particular
complex structure labelled by a single $\tau$ with its expression
obtained in a straightforward manner from the general formula of
Eq.(\ref{tauexpl}) or directly from the corresponding formula of
the STU truncation. For this truncation and special complex
structure the "bosonic" three-qubit state is of the usual form
\beq \vert\Gamma\rangle=(-{ d})^{1/4}\left(e^{i\alpha}\vert
000\rangle -e^{-i\alpha}\vert\overline{111}\rangle\right),\qquad
\tan\alpha=\frac{p^0}{\hat{p}^0}. \label{GHZattrt3} \eeq \noindent
where now $d$ is given by the expression of
Eq.(\ref{Cayleygen22}). As usual the charge states supporting BPS
black holes (among other conditions\cite{stu}) have $d<0$. The
semiclassical black hole entropy is given by \beq S_{BH}=\pi
V_H(\varrho_{crit})=\pi\sqrt{- d}. \label{entrasameasure2} \eeq
\noindent

\section{Generalized Hitchin Functional}

\subsection{Quantum corrections}
We have seen in the previous section that Hitchin's functional
$V_H(\varrho)$ at its critical point has a very important physical
interpretation. According to Eq.(\ref{lenyeg}) it is just
proportional to the semiclassical black hole entropy $S_{BH}$ which can also be related to an entanglement measure. Now
we will regard $S_{BH}$ merely as the leading-order contribution to the
black hole entropy. In order to do this one can consider the
quantum theory with action $V_H$. Let us formally define the
partition function\cite{Dijkgraaf} \beq
Z_H(\gamma)=\int_{[\varrho]=\Gamma} e^{V_H(\varrho)+d\sigma}{\cal
D}\sigma \label{partition} \eeq \noindent where as usual $\Gamma$
is the Poincar\'e dual to $\gamma$.
Then in Ref.\cite{Dijkgraaf}
it has been conjectured that the partition function $Z_H(\gamma)$ on a
manifold $M$ is the Wigner transform of the partition functions of
$B$ and $\overline{B}$ topological strings on $M$. Based on our considerations of Section 3.2 using the method of steepest descent it is easy to demonstrate
that this conjecture is correct at the classical level. Then the
idea was to use Hitchin's functional also to recover the quantum
corrections that has already been calculated via topological
string techniques\cite{Cecotti}. However, it turned out\cite{PW}
that after appropriate gauge fixing at the one loop level there is
a discrepancy between the result based on Hitchin's functional and
the result of topological string theory. In order to resolve this
discrepancy Pestun and Witten suggested to use a partition
function based on the {\it generalized Hitchin functional}
instead. Hitchin's functional is connected to Calabi-Yau
structures on the other hand the generalized Hitchin functional
(GHF) is connected to generalized Calabi-Yau
structures\cite{genHitchin,Gualtieri}. For the resolution they
have chosen manifolds with $ b_1(M)=0$ where the critical points
and classical values of both functionals coincide, however the
quantum fluctuating degrees of the two functionals are different.
The upshot of these consideration was that after a convenient
interpretation\cite{PW} the conjecture of
Ref.\cite{Dijkgraaf} remains true even at the one loop level.

Hence we have an interesting possibility of using form theories of
gravity in a much wider context which is also capable of
 describing quantum fluctuations provided we are willing to use
a generalization of Hitchin's functional. Within the framework of
the BHQC we have seen that Hitchin's functional is related to an
entanglement measure and its descendants that can describe
entangled systems with a small number of constituents. Now the
question is whether we can also find a quantum information
theoretic interpretation of the GHF. The aim of this section is to
show that a variant of the entangled system which is directly
connected to the invariant underlying the GHF has already been
discussed in the BHQC\cite{Duff2,Levayfano}. This entangled system
lends itself to precisely such an interpretation. Hence as
an extra bonus its intimate connection to the GHF enables a
further generalization of the BHQC.

The GHF for a six dimensional manifold $M$
is defined by replacing the three-form $\varrho_3$ in the usual formulation of the Hitchin functional by a {\it polyform}  $\varphi=\varphi_1+\varphi_3+\varphi_5$ of odd degree. It was shown\cite{genHitchin} that if this polyform is {\it nondegenerate} in a suitable sense then it defines a generalized almost complex structure\cite{genHitchin} on $M$.
The nondegeneracy is defined via a quartic invariant, our main concern here,
which is invariant under $Spin(6,6)$.
Then the generalized complex structure is given by a pure
spinor\cite{genHitchin,Gualtieri,Chevalley} of the
form $\varphi+i\hat{\varphi}(\varphi)$ with respect to
$Spin(TM,TM^{\ast})$ where $TM$ and $TM^{\ast}$ are the tangent and cotangent bundles of $M$. It is then shown that if in addition $d\varphi=0$ and $d\hat{\varphi}=0$ then the generalized almost complex structure is integrable giving rise to a {\it generalized complex manifold}.

As far as physics is concerned the interest in such manifolds stems from the fact that a special case of such manifolds is the class of {\it generalized Calabi-Yau} manifolds. Such manifolds are showing up in strings propagating in general backgrounds. A generalized Calabi-Yau structure (GCYS) can be regarded as a one which is interpolating in a suitable sense between the symplectic structure and the Calabi-Yau one of a given manifold $M$.
Similarly to the Hitchin functional case the condition which is crucial for the GCYS namely $d\hat{\varphi}=0$ is arising from the extremization of the GHF which is constructed from the quartic invariant.
There is an alternative formulation of the GHF
based on polyforms of even rank. These polyforms are of the form
$\varphi=\varphi_0+\varphi_2+\varphi_2+\varphi_6$.
Note that the dimension of the space of such polyforms is $32$ in both cases.
This is connected to the fact that polyforms with odd (even) degree form  an irrep of negative (positive) chirality with respect to $Spin(6,6)$.
Our aim is to present a form for the quartic invariant related to the GHF
based on the Freudenthal system corresponding to the Jordan algebra of quaternion Hermitian $3\times 3$ matrices.
Note that the usual form of the quartic invariant underlying the GHF is based onthe moment map and the properties of pure spinors\cite{genHitchin}.
This alternative form given below is convenient for our entanglement based considerations.
In the Appendix for the convenience of the reader we have collected the mathematical results on polyforms and the Freudenthal system
we need for our presentation.

Recall first that after complexification ${\cal D}$ of Eq.(\ref{Cayleygen})
underlying the construction of Hitchins functional
was a relative invariant under $G\equiv GL(6,{\mathbb C})=GL(1,{\mathbb C})\times SL(6,{\mathbb C})$. Moreover, this relative invariant
was related to the PV of Sato and Kimura of Class 5.
We also used a suitable restriction of this relative invariant for the description of embedded entangled systems. The restricted Hitchin functional in this case
was based on the group $G\equiv GL(1,{\mathbb C})\times Sp(6,{\mathbb C})$.
This relative invariant was related to the PV of Class 14.
We also observed that our relative invariants could elegantly be described by
Freudenthal systems based on the {\it complexifications} of the Jordan algebras
of $3\times3$  complex Hermitian or real symmetric matrices.
The crucial identity in this respect was the complex analogue of Eq.(\ref{Hamiltoni1}).
\beq
2\sqrt{\vert{\cal D}(P)\vert}e^{i{\rm arg}{\cal D}(P)}=\{P,\hat{P}\}
\label{crucial}
\eeq
\noindent
where $P\in \wedge^3 {\mathbb C}^6$ or
$P\in \wedge^3_0 {\mathbb C}^6$ respectively and $\hat{P}$ was the corresponding complex extension of the Freudenthal dual.
Nondegenerate three-forms then corresponded to three-fermion states belonging to the {\it complex} GHZ class characterized by the property ${\cal D}\neq 0$.
This class was then a {\it stable orbit} under $G$.
Moreover, the relative invariant ${\cal D}$ and the product $\{\cdot,\cdot\}$ was just the {\it negative} of the quartic invariant and the symplectic form of the corresponding Freudenthal system.

We have seen that over the reals the complex GHZ class splits into two real GHZ orbits with the corresponding states having ${\cal D}<0$ and ${\cal D}>0$ respectively.
For the construction of Calabi-Yau structures the first of the {\it real orbits} was needed.
As we already know  a real state belonging to this orbit
can be exressed as a real part of a complex separable state. On this open orbit the real SLOCC group $GL(1,{\mathbb R})\times SL(6,{\mathbb R})$ is acted transitively.
For the generalized Calabi-Yau structures it is known\cite{genHitchin} that
there is a real spinor
which is the real part of a complex {\it pure} spinor.
This spinor is
belonging to the $32$ dimensional spinor representation with the corresponding quartic invariant being negative.
Moreover, spinors with this property form an open set and the real group $GL(1,{\mathbb R})\times Spin(6,6)$ is acting on it transitively.
More importantly this open orbit can be regarded as one of the two real orbits arising from the stable complex orbit of a PV with group $G=GL(1,{\mathbb C})\times Spin(12,{\mathbb C})$  which is Class 23 in the Sato-Kimura scheme\cite{Satokimura}.
This is as we expected the next item in the line of Freudenthal systems, namely the one which is based on the complexification of the Jordan algebra of $3\times3$ quaternion Hermitian matrices.

The upshot of these investigations is that a nice alternative way for expressing the GHF is simply using the real
version of Eq.(\ref{crucial}) with ${\cal D}<0$ with now ${\cal D}$ is 
replaced by a ${\mathbb D}$ which is the negative of the quartic invariant of the corresponding Freudenthal system.
What is left to be established is the precise dictionary between the components of a polyform $\varphi$ and the components of the Freudenthal system.
A detailed derivation of this is given in the Appendix.
Choosing a polyform of even degree $\varphi=\varphi_0+\varphi_2+\varphi_4+\varphi_6$ the GHF is
\beq
V_{GH}(\varphi)=\int_M\sqrt{-{\mathbb D}(\varphi)}d^6x=-\frac{1}{2}\int_M\varphi\wedge\hat{\varphi}(\varphi)
\label{genhitfunk}
\eeq
\noindent
where
\beq
{\mathbb D}(\varphi)=-q(p),
\label{kapcsolat}
\eeq
\noindent
is the quartic invariant of $p$ which is the element of the Freudenthal system.
For the explicit form of $q(p)$ and the correspondence $\varphi\leftrightarrow p$ see Eqs.(\ref{quart2})-(\ref{masod}) of the Appendix.
Notice that for the nonlinear expression $\hat{\varphi}(\varphi)$ now we have an explicit form in terms of the Freudenthal dual of the Freudenthal system.
For explicit formulae one just has to use the quaternionic analogues of Eqs.(\ref{tild1})-(\ref{tild2}).
The Freudenthal formalism again automatically takes care of the nice symplectic
 interpretation. Namely the space of polyforms can be regarded as a phase space of a classical mechanical system. The symplectic form is given by the pairing $\{\cdot,\cdot\}$ and $\sqrt{\vert{\mathbb D}(\varphi)\vert}$ is the Hamiltonian.
The Freudenthal dual $\hat{\varphi}$ is up to sign just the Hamiltonian vector field. The generalized almost complex structure is integrable for $d\hat{\varphi}=0$.

\subsection{The generalized Hitchin invariant as an entanglement measure}

Interestingly an entangled system that we can relate to the invariant underlying the GHF has already appeared in the literature of the BHQC\cite{Duff2,Levayfano}.
In order to see this notice that
there is yet another PV that we have not discussed yet.
It is Class 29. in the Sato-Kimura list. The group of the PV in this case is $GL(1,{\mathbb C})\times E_7({\mathbb C})$.
The corresponding Freudenthal system is the one based on the complexification of the Jordan algebra of $3\times 3$ octonion Hermitian matrices.
The quartic invariant of the associated Freudenthal system is well-known
in the string theory literature.
Indeed the most general class of black holes in
$\mathcal{N}=8$ supergravity/M-theory
is defined by $56$ charges and the entropy formula is
given by the square root of
the quartic Cartan-Cremmer-Julia $E_{7(7)}$ invariant  \cite{Cartan,Cremmer,Kol}
which is a real version of our quartic invariant.

It can be shown that the $56$ dimensional fundamental representation of $E_7$ can be decomposed with respect to the $SL(2,{\mathbb C})^{\times 7 }$ subgroup as follows\cite{Duff2,Levayfano,cherchiai}
\begin{eqnarray}
\label{dek}
{\bf 56}&\to& ({\bf 2},{\bf 2},{\bf 1},
{\bf 2},{\bf 1},{\bf 1},{\bf 1})+
({\bf 1},{\bf 2},{\bf 2},{\bf 1},{\bf 2},{\bf 1},{\bf 1})+
({\bf 1},{\bf 1},{\bf 2},{\bf 2},{\bf 1},{\bf 2},{\bf 1})\nonumber\\&+&
({\bf 1},{\bf 1},{\bf 1},{\bf 2},{\bf 2},{\bf 1},{\bf 2})+
({\bf 2},{\bf 1},{\bf 1},{\bf 1},{\bf 2},{\bf 2},{\bf 1})\\&+&
({\bf 1},{\bf 2},{\bf 1},{\bf 1},{\bf 1},{\bf 2},{\bf 2})+
({\bf 2},{\bf 1},{\bf2},{\bf 1},{\bf 1},{\bf 1},{\bf 2})\nonumber.
\end{eqnarray}
\noindent
Let us now replace formally the $2$s with $1$s,
and the $1$s with $0$s, and form a $7\times 7$
 matrix by regarding the seven vectors obtained in this way as its rows.
 Let the rows correspond to lines and the
 columns to points, and the location of a ``1'' in the
 corresponding slot correspond to incidence.
 Then this correspondence results in the incidence matrix of the Fano plane
Let us reproduce here this incidence matrix with the following labelling for the
 rows (r) and columns (c)
\beq
\begin{pmatrix}r/c&A&B&C&D&E&F&G\\
               a&1&1&0&1&0&0&0\\
                b&0&1&1&0&1&0&0\\
              c&0&0&1&1&0&1&0\\
                   d&0&0&0&1&1&0&1\\
                  e&1&0&0&0&1&1&0\\
                f&0&1&0&0&0&1&1\\
               g&1&0&1&0&0&0&1\end{pmatrix}\mapsto
  \begin{pmatrix}a_{ABD}\\ b_{BCE}\\c_{CDF}\\d_{DEG}\\e_{EFA}\\f_{FGB}\\g_{GAC}
           \end{pmatrix}
       \label{konvencio1}
       \eeq
   \noindent
where we also displayed the important fact that this labelling automatically
defines
the index structure for the amplitudes
of seven three-qubit states formed
out of seven distinguishable qubits $A,B,C,D,E,F,G$
. If we introduce the notation $V_{ijk}\equiv V_i\otimes
   V_j\otimes V_k$ where $i,j,k\in \{A,B,C,D,E,F,G\}$ then the ${\bf
   56}$ of $E_7$ denoted by ${\cal H}$ decomposes as \beq {\cal
   H}=V_{ABD}\oplus V_{BCE}\oplus V_{CDF}\oplus V_{DEG}\oplus
   V_{EFA}\oplus V_{FGB}\oplus V_{GAC}. \label{decompcycle} \eeq
   \noindent Clearly this structure encompasses an unusual type of
   entanglement. For entanglement is usually associated with tensor
   products, however here we also encounter {\it direct sums}. One
   can regard the seven tripartite sectors as seven superselection
   sectors which in the black hole context corresponding to seven different $STU$
   truncations\cite{Duff2,Levayfano}. This structure  is usually
   referred to in the literature as the tripartite entanglement of
   seven qubits\cite{Duff2}.

Now one can express the quartic invariant in terms of the $56$ amplitudes
of the seven copies of three-qubit systems\cite{Duff2,Levayfano,review2,cherchiai}.
Especially one can establish a precise dictionary between this qubit based
description and the Freudenthal one\cite{review2}.
Then one can consider
the decomposition of the ${\bf 56}$ of $E_7({\mathbb C})$
\beq
{\bf 56}\to ({\bf 2},{\bf 12})\oplus ({\bf 1}, {\bf 32})
\eeq
\noindent
with respect to the subgroup $SL(2,{\mathbb C})\times Spin(12,{\mathbb C})$.
It can then be shown that the $({\bf 1},{\bf 32})$ part is consisting of those amplitudes that are excluding one particular qubit.
Hence for example the space
\beq V_{BCE}\oplus V_{CDF}\oplus V_{DEG}\oplus V_{FGB}
\eeq
\noindent
excluding qubit $A$ form a representation space for $Spin(12,{\mathbb C})$.
This representation space comprises the tripartite entanglement of ${\it six}$
qubits. Now after using the correspondence between the $32$ amplitudes
$b_{BCE},c_{CDF},d_{DEG}$ and $f_{FGB}$ and the $1\oplus 15\oplus 15\oplus 1$
structure of the relevant Freudenthal system one can see that the quartic invariant is just the same as the one underlying the GHF.

As far as string theory is concerned this entanglement based interpretation is useful because it reveals four STU subsectors hidden in the structure of the GHF. However there is an even more useful interpretation.
It is just the one of directly regarding polyforms as representatives of fermionic entangled systems and the relative invariant like the one underlying the GHF as an entanglement measure.
Of course since the structure of polyforms is also intimately connected to the structure of the underlying manifold $M$ and its moduli space this interpretation is again an unusual one. Recall also that the number of modes or single particle states
is just equals the dimension of $M$.

Notice moreover that the fermion number is not conserved.
For our case of the GHF we have either $\varphi=\varphi_1+\varphi_3+\varphi_5$
or $\varphi=\varphi_0+\varphi_2+\varphi_4+\varphi_6$.
The analogue of the SLOCC group is now $GL(1,{\mathbb C})\times Spin(12,{\mathbb C})$ which is mixing the forms of different degree but respecting the parity.
One can write these polyforms as a sum of objects like in Eq.(\ref{itt1}) expressed in terms of different numbers of fermionic creation operators.
Alternatively one can regard the polyforms as spinors\cite{genHitchin,Gualtieri} .
It is easy to see that {\it pure spinors} should correspond in this picture to separable states.
Classifying the entanglement types of spinors then should correspond to finding the SLOCC orbits and the stabilizers of the representatives. This problem has been solved up to dimension twelwe in the classical paper of Igusa\cite{Igusa}. Many results can also be found in the book of Chevalley\cite{Chevalley}.

Notice that the $B$-transform of the pure spinor $1$ (i.e. a ${\varphi}_0$), $e^{-B}\cdot 1$ (another pure spinor) can be expressed in terms of the Pfaffian combinations
of the $6\times 6$ matrix ${\cal B}$ underlying the two-form $B$.
(See Eq.(\ref{naezaz}))
There is a similar phenomenon occurring
in the theory of fermionic Gaussian states where the higher order correlations can be obtained from the quadratic ones via Wick's theorem.
(See for example Eqs. (4) and (8) of the paper of Kraus et.al.\cite{Kraus}).
This can yield a standard form for Gaussian states that look like "paired states" known from the BCS theory of superconductivity\cite{Kraus,Botero}.
These states look similar to GHZ-states coming from combinations of two pure spinors.
There are many more mathematical correspondences with fermionic systems in many body physics which deserve some attention. However, as far as the authors are aware this perspective have not made its debut to the Quantum Information community.
The elaboration of these ideas within the field of quantum entanglement could be another useful input string theory can provide.

\subsection{An example. $T^6$ revisited.}

A simple example illustrating the difference between the Hitchin and generalized Hitchin functionals was given by Pestun\cite{Pestun}.
Here we briefly revisit this example putting the emphasis on the entanglement interpretation.

We already know that for Calabi-Yau compactification in the supergravity approximation black hole entropy is equal to Hitchin functional taken at its critical point. We have also seen that at the critical point the resulting expression can be interpreted as an entanglement measure.
Moreover, according to the OSV relation\cite{OSV} black hole entropy can be related to the topological string partition function.
Hence in terms of partition functions at the classical level we have checked the chain
of relations symbolically written as $Z_H=Z_{BH}=\vert Z_{TOP}\vert^2$.
For Calabi-Yau manifolds $M$ with $b_1(M)=0$ in order to have these relations even at at one loop level
we have learnt that we have to replace $Z_H$ with $Z_{GH}$ where the latter is the partition function based on the GHF.
Can we relate the GHF taken at its critical point to black hole entropy as an entanglement measure already at the tree-level?  Clearly to have this situation we need a manifold where $b_1(M)\neq 0$.
The simplest example of that kind\cite{Pestun} is $T^6$.
This is of course our example already used in connection with the Hitchin functional.
However, now we will suppose that the extra fields featuring the GHF have nonzero expectation values even at tree level.
Note that although this illustrative case has more than $N=2$ supersymmetry (the setting needed for topologiclal strings) hence we do not expect
agreement at the one loop level,
at the tree level it still has a consistent $N=2$ truncation.

In this $T^6$ example the $N=8$ supergravity multiplet is truncated by disregarding the gravitini multiplet with a result of having instead of the $1+12+15$ gauge fields the $1+15$ ones of the $N=2$ sector.
As a result of this we are merely having those vector multiplets at our disposal whose corresponding scalars are giving rise to the generalized complex moduli of $T^6$.
In the case of the IIA picture telated to the topological A model
the 32 charges correspond to the wrapping configurations of the $D0,D2,D4,D6$ branes.
The periods $X^I, I=0,1,\dots 15$  are arising from integrals of the komplexified K\"ahler class $\Omega$
\beq
\Omega=e^{b+i\omega}=\varphi+i\hat{\varphi}(\varphi),\qquad \varphi=\varphi_0+
\varphi_2+\varphi_4+\varphi_6.
\label{kahlerclass}
\eeq
\noindent
where now
\beq
\varphi=[\Gamma],\qquad \Gamma=p^0\alpha_0+P^{\mu\nu}\alpha_{\mu\nu}-Q_{\mu\nu}
\beta^{\mu\nu}-q_0\beta^0.
\eeq
\noindent
Here $\alpha_0, \alpha_{\mu\nu}, \beta^{\mu\nu}, \mu,\nu=1,2,\dots 6$ and $\beta^0$ are a basis
of $0,2,4$ and $6$-forms.
Now an expansion similar to the one as given by Eq.(\ref{kifejtes2})
gives for the complexified Kahler class $\tau^{\mu\nu}=X^{\mu\nu}/X^0$
corresponding to a critical point of the GHF
with an expression for the generalized complex structure as given by Eq.(\ref{tauexpl}).
Note that this formula is the same in appearance, however now the matrices
are $6\times 6$ antisymmetric ones or $3\times 3$ ones with biquaternionic entries.
This is according to the Appendix just the complexification of the Freudenthal system based on the Jordan algebra of $3\times 3$ quaternion Hermitian matrices.
For an explicit mapping between these matrices see Eq.(\ref{identifi}).
Note that in this formalism $\hat{\varphi}$ is again the Freudenthal dual as determined by the cohomology class $[\Gamma]$ of the charge polyform.
Now after performing manipulations and using identities for the corresponding Freudenthal system the GHF at the critical point can be evaluated.
The result for the semiclassical black hole entropy is as expected\cite{Pioline,Pestun}
\beq
S_{BH}=\pi V_{GH}(\varphi_{crit})=\pi\sqrt{-\mathbb{D}}
\eeq
\noindent
where ${\mathbb{D}}$ is related to the quartic invariant of the quaternionic Freudenthal system of Eq.(\ref{quart2})  as
\beq {\mathbb{D}}(p^0,q_0,P,Q)=-q(\xi,\eta,X,Y).
\eeq
\noindent
According to the results of the previous subsection we can reinterpret this formula as an entanglement measure describing the tripartite entanglement
of six qubits, or of a fermionic system with six modes and even parity.
Notice also that the distillation interpretation of the "attractor states" of
Eq.(\ref{GHZattr}) also holds in this case, with the basis "states" like $\vert 123\rangle$ and $\vert\overline{123}\rangle$, ${\cal D}$ and $\hat{p^0}$ should be replaced with the corresponding
pure spinors, ${\mathbb D}$ and the relevant component of the Freudenthal dual.

\section{Entanglement of three fermions
 with seven single particle states}

In this section as a further step we would like to propose a
reinterpretation of the invariant underlying Hitchins functional
for three-forms in seven dimensions as an entanglement measure for
three fermions with seven single particle states. As a byproduct
of this we can simply use the classification
theorem\cite{Schouten,Cohen} of $GL(7,{\mathbb C})$ orbits of
three forms in seven dimensions to give a full list of SLOCC
entanglement classes. Note that this invariant integrated on a
real seven manifold is well-known in string theory. The critical
points of the associated functional give rise to manifolds with
$G_2$ holonomy. At the critical point the nondegenerate
three-forms not only determine a metric of $G_2$ holonomy but also
special three dimensional submanifolds with minimal volume.
The interpretation of this new functional as a one related to another entanglement measure provides a
further support for the BHQC.

Let us denote the octonionic units as $e_1, e_2,\dots e_7$,
and for their multiplication table use the conventions of G\"unaydin and G\"ursey\cite{Gunaydin}.
With this notation an octonion $x\in {\mathbb O}$ and
its conjugate  $\overline{x}$ can be written as
$x=x_0+x_Ae_A$  and $\overline{x}=x_0-x_Ae_A$ where summation for $A=1,2,\dots 7$ is implied.
An imaginary octonion $x=x_Ae_A$
in the basis of $e_A$ has the usual norm $Q(x)=x\overline{x}=x_1^2+\dots x_7^2$.

Let us now consider the seven dimensional complex vector space
$U$. By an abuse of notation we also denote its canonical basis
vectors by $e_A$. Let us denote the six dimensional subspace $V$
of $U$ spanned by $e_a, a=1,\dots 6$. The basis vectors for the
dual $U^{\ast}$ will be denoted by $e^A$. As a complex basis of
$U^{\ast}$ we define \beq E^{1,2,3}=e^{1,2,3}+ie^{4,5,6},\qquad
E^{\overline{1},\overline{2},\overline{3}}=e^{1,2,3}-ie^{4,5,6},
\qquad E^7=ie^7. \label{ebasis} \eeq \noindent Let us use in the
following the shorthand notation $e^{ABC}\equiv e^A\wedge
e^B\wedge e^C$. For $1\leq A<B<C\leq 7$ the $e^{ABC}$ form a basis
for $\wedge^3U^{\ast}$. Then a $GHZ$-like state in the subspace
$\wedge^3V^{\ast}$ can be written as \beq E^{123}
+E^{\overline{1}\overline{2} \overline{3}}=2(e^{123}-e^{156}
+e^{246}-e^{345}) \eeq \noindent With the usual relabelling $4, 5,
6\mapsto \overline{1},\overline{2},\overline{3}$ and up to
normalization the state on the right hand side is just the one of
Eq.(\ref{Dnegativ}) with its Hitchin invariant of
Eq.(\ref{Cayleygen})  being negative. Let us add to this state the
one $(E^{1\overline{1}}+ E^{2
\overline{2}}+E^{3\overline{3}})\wedge E^7$. Then we obtain the
three fermion state with seven single particle states \beq
{\phi}\equiv
\frac{1}{2}(E^{123}+E^{\overline{123}}+(E^{1\overline{1}}
+E^{2\overline{2}}+E^{3\overline{3}})\wedge
E^7)=e^{123}-e^{156}+e^{246}-e^{345} +e^{147}+e^{257}+e^{367}.
\label{calibration} \eeq \noindent Notice that the structure of
our tripartite state ${\phi}$ is encoded into the incidence
structure of the {\it lines} of the {\it oriented} Fano plane
which is also encoding the multiplication table of the
octonions\cite{Gunaydin}. As a {\it complex} three-form it can be
shown\cite{Schouten,Bryant} that the subgroup of the SLOCC group
$GL(7,{\mathbb C})$ that fixes $\phi$ is $G^{\mathbb
C}_2\times\{\omega{\bf 1}\vert\omega^3=1\}$ where ${\bf 1}$ is the
$7\times 7$ identity matrix.

Rather than using $\phi$ as an entangled state, in
string theory it is used as a {\it real} differential form on a seven
dimensional real manifold.
In this context instead of the complex SLOCC group the real one i.e. $GL(7,{\mathbb R})$ is used.
The stabilizer of $\phi$ as a real three-form is the compact real form $G_2$
which is the automorphism group of the octonions. In the theory of special holonomy
manifolds invariant forms like $\phi$ are called as calibrations.
Note that after the permutation $e^5\leftrightarrow e^7$ we obtain
the form for $\phi$ usually used in the
literature\cite{Becker,Bryant}.

Let us now take an {\it arbitrary} element $\Phi\in
\wedge^3U^{\ast}$ with $U={\mathbb C}^7$. Such an element can be
written in the form \beq \Phi=\frac{1}{3!}\Phi_{ABC}e^A\wedge
e^B\wedge e^C. \eeq \noindent Define\cite{Hitchin,Dijkgraaf} the
matrix of a symmetric bilinear form as \beq {\cal
B}_{AB}=-\frac{1}{144}\Phi_{AC_1C_2}\Phi_{BC_3C_4}\Phi_{C_5C_6C_7}
\epsilon^{C_1C_2C_3C_4C_5C_6C_7}.\label{metrikahoz1} \eeq
\noindent Then it can be shown\cite{Satokimura,Hitchin} that \beq
I_7({\Phi})\equiv {\rm Det}{\cal B} \label{I7} \eeq \noindent is a
relative invariant under the action of the SLOCC group
$GL(7,{\mathbb C})$. This means that under the action of a $g\in
GL(7,{\mathbb C})$ the invariant transforms as $g^{\ast}I_7=({\rm
Det}g)^9I_7$. Especially choosing $\phi$ shows that ${\cal
B}_{AB}=\delta_{AB}$ hence $I_7(\phi)=1$.

We propose the three fermionic state $\phi$ of
Eq.(\ref{calibration}) as a generalization of the tripartite GHZ
state. It has a nonvanishing relative invariant just like the GHZ
state for three qubits, and the GHZ-like states for three fermions
with six single particle states. The invariant $I_7$ for three
fermionic states plays a similar role than Cayley's
hyperdeterminant Eq.(\ref{Cayley}) for three qubits. There is
another similarity with the canonical GHZ state and $\phi$. If we
suitably normalize $\phi$ hence producing a $\vert\phi\rangle$
with unit norm and calculate the reduced density matrix (since the
constituents are identical any of such reduced density matrices
will do) we get \beq \rho_1\equiv {\rm
Tr}_{23}\vert\phi\rangle\langle \phi\vert=\frac{1}{7}{\bf 1}.
\label{reduced1} \eeq \noindent This reduced density matrix is the
one representing the {\it totally mixed state} for any of the
subsystems. This relation is coming from the identity
$f_{ACD}f_{BCD}=6{\delta}_{AB}$ for the octonionic structure
constants i.e. $e_Ae_B=f_{ABC}e_C$. The two-partite reduced
density matrix of $\vert\phi\rangle$ \beq \rho_{23}={\rm
Tr}_1\vert\phi\rangle\langle\phi\vert \eeq \noindent will be a
$21\times 21$ matrix. The structure of this matrix can be worked
out using the identity \beq
f_{ABC}f_{ADE}=f_{BCDE}+{\delta}_{BD}{\delta}_{CE}-{\delta}_{BE}{\delta}_{CD}.
\eeq \noindent This formula shows that the structure of bipartite
density matrices is controlled by the octonionic structure
constants $f_{BCDE}$ connected to the incidence structure of the
complement of the lines of the Fano plane. Hence regarded as an
entangled state $\vert\phi\rangle$ is connected in many ways to
the structure of the octonions. It would be an interesting
possibility to to use the properties of $\vert\phi\rangle$ as a
manifestation of the algebra of octonions in quantum information.

In string theory instead of the complex vector space $U$ the real
tangent space of a seven manifold $M_7$ is used. This can be
regarded as the real version of the state space for our tripartite
states with the amplitudes now depending on the coordinates of the
manifold. Hence the state in this case is a {\it real} differential
three-form. For nondegenerate three forms ${\Phi}$ taken from the stable orbit of $GL(7,{\mathbb R})$
represented by $\phi$ one can define a
metric\cite{Hitchin,Dijkgraaf} \beq g_{AB}={\rm Det}({\cal
B})^{-1/9}{\cal B}_{AB}. \label{metrikahoz2}\eeq \noindent Since
${\rm Det}g=({\rm Det}{\cal B})^{2/9}$ one can define Hitchin's
functional \beq
V_7(\Phi)\equiv\int_{M_7}I_7^{1/9}(\Phi)d^7x=\int_{M_7}
\sqrt{g_{\Phi}}d^7x. \label{Hitchin7} \eeq \noindent This formula
shows that Hitchin's functional is simply the volume of $M_7$ with
respect to a metric determined by the nondegenerate three-form
$\Phi$
according to the formulas of
Eqs.(\ref{metrikahoz1}) and (\ref{metrikahoz2}).
The relative
invariant $I_7$ is just the entanglement measure of Eq.(\ref{I7})
we have discussed above. The important property of $V_7(\Phi)$ is
that its critical points in a fixed cohomology class
give\cite{Hitchin,Dijkgraaf} \beq d\Phi=0,\qquad d\ast\Phi=0
\label{g2eq} \eeq \noindent where the Hodge star is the one
defined with respect to the metric determined by $\Phi$. These are
the conditions for our three-form $\Phi$ defining a metric of
$G_2$ holonomy\cite{Bryant}.

\begin{table}[h]
\centering
\begin{tabular}{|c|c|c|c|}
\hline
Type & Canonical form & Instructive $SL(7)$ equivalent & Name \\ \hline \hline
$f_1$ & $E^{123}$ & $E^{123}$ & Sep\\
$f_2$ & $E^{123}+E^{145}$ & $E^1\wedge (E^{23}+E^{\bar 2\bar 3})$ & Bisep\\
$f_3$ & $E^{123}+E^{456}$ & $E^{123} +E^{\bar 1\bar 2\bar 3}$ & GHZ\\
$f_4$ & $E^{162}+E^{243}+E^{135}$& $E^{12\bar 3}+E^{1\bar 2 3} + E^{\bar 1 2 3}$
 & W \\
 $f_5$ & $E^{123}+E^{456}+E^{147}$ & $E^{1\bar 1}\wedge E^7+E^{123} +
 E^{\bar 1\bar
  2\bar 3}$ & Sympl$_1$/GHZ\\
 $f_6$ & $E^{152}+E^{174}+E^{163}+E^{243}$ & $(E^{1\bar 1}+E^{2\bar 2}+E^{3\bar 3
 })\wedge E^7 +E^{\bar 1\bar 2\bar 3}$ & Sympl$_3$/Sep \\
 $f_7$ & $E^{146}+E^{157}+E^{245}+E^{367}$ & $(E^{2\bar 2}+E^{3\bar 3})\wedge E^7
  +E^{123} +E^{\bar 1\bar 2\bar 3}$ & Sympl$_2$/GHZ \\
  $f_8$ & $E^{123}+E^{145}+E^{167}$ & $(E^{1\bar 1}+E^{2\bar 2}+E^{3\bar 3})\wedge
   E^7$ & Sympl$_3$ \\
   $f_9$ & $E^{123}+E^{456}+(E^{14}+E^{25}+E^{36})\wedge E^{7}$ & $(E^{1\bar 1}+E^{
   2\bar 2}+E^{3\bar 3})\wedge E^7+E^{123} +E^{\bar 1\bar 2\bar 3}$ & Sympl$_3$/GHZ
   \\  \hline
   \end{tabular}
\caption{Entanglement classes of three fermions with seven single particle states.}
\label{tab:1}
\end{table}

Apart from the nondegenerate class (i.e. the one with $I_7\neq 0$)
in quantum information one is also interested in the full
structure of $GL(7,{\mathbb C})$ orbits and their stabilizers.
These classes are precisely the SLOCC entanglement classes. The
orbit structure over the complex field has been given by
Schouten\cite{Schouten} over finite fields it has been obtained by
Cohen and Helminck\cite{Cohen}. Here we need the result over
${\mathbb C}$. In the notation of Ref.\cite{Cohen} these are the
classes of type $f_1-f_9$ see the first column of Table I. Here in
accordance with the notation of Ref.\cite{Cohen} in the second
column we expressed the representatives of these classes in the
basis $\{E^A\}$. Again it is
instructive to relabel the basis vectors using the mapping
$\{1,2,3,4,5,6,7\}\mapsto
\{1,2,3,\overline{1},\overline{2},\overline{3},7\}$. In this new
notation the representatives of the SLOCC classes are given in the
third column of Table I. Note that arriving at these forms for the
classes $f_2,f_6,f_7,f_8$ we have chosen different representatives
by applying suitable permutations that are still elements of the
SLOCC group. The corresponding permutations are: $(456)$,
$(17346)(25)$, $(1765342)$, $(176342)$ respectively.

Notice that in the second column of Table I. the canonical forms
are written in the form of $\varrho+\omega\wedge E^7$ where
$\varrho$ is a three-form based on the six dimensional subspace
$V^{\ast}$ spanned by the basis vectors $E^a,\quad a=1,\dots 6$
and $\omega$ is either zero, or a two form on $V$ of
Slater\cite{Schlie} rank $1, 2$ or $3$. For $\omega\equiv 0$ the
three-form $\varrho$ can belong to the four classes well-known
from Section 2.2.  They are the separable, biseparable, W and GHZ
states. There is a class with $\varrho\equiv 0$ i.e. our state is
of the form $\omega\wedge E_7$ with $\omega$ a nondegenerate
symplectic form. We have three classes with $\varrho$ being a GHZ
state combined with a term $\omega\wedge E_7$ with the two
form being Slater rank $1,2,3$. Notice that the case of maximal
Slater rank plus a GHZ state is just the nondegenerate state
$\phi$ belonging to the complex stable orbit of $GL(7,{\mathbb
C})$. There is still one class we have not mentioned, it is the
one with a representative consisting of a fully separable
$\varrho$ plus $\omega\wedge E_7$ with $\omega$ full rank.

It is important to note that over the reals we have {\it two}
stable $GL(7,{\mathbb R})$ orbits. One of them is just the one
with the usual representative $\phi$ of Eq.(\ref{calibration})
expressed in the real basis $e_A$. Its stabilizer is the compact
real from $G_2$ of the complex group $G^{\mathbb C}_2$. The other orbit has the
representative
 \beq
\tilde{\phi}=e^{123}+e^{345}+e^{156}-e^{246}-e^{147}-e^{257}-e^{367}
\label{splitcalibration} \eeq \noindent with its stabilizer being
$\tilde{G}_2$ the noncompact real form of $G^{\mathbb C}_2$, i.e. the automorphism
group of the split octonions. Using the new basis \beq
F^{1,2,3}=e^{1,2,3}+e^{4,5,6},\qquad
F^{\overline{1},\overline{2},\overline{3}}=e^{1,2,3}-e^{4,5,6},\qquad
F_7=e^7 \label{fbasis}\eeq
 \noindent
$\tilde{\phi}$ can be written as \beq
\tilde{\phi}=\frac{1}{2}(F^{123}+F^{\overline{123}}+(F^{1\overline{1}}+F^{2\overline{2}}+
F^{3\overline{3}})\wedge F^7). \label{fitilda}\eeq \noindent
Comparing Eqs.(\ref{calibration}) and (\ref{fitilda}) we see that $\phi$ and $\tilde{\phi}$ are of the same form in the basis $\{E^A\}$ and $\{F^A\}$ respectively.
From the definitions of Eq.(\ref{ebasis}) and (\ref{fbasis}) it is clear that although $\phi$ and $\tilde{\phi}$ are $GL(7,{\mathbb R})$ inequivalent but they are
$GL(7,{\mathbb C})$ equivalent.
Observe that $\phi$ and $\tilde{\phi}$ can be written in the canonical form $\varrho_{\mp}\pm\omega\wedge e^7$ where $\varrho_{\mp}$ are
three-forms with Hitchins invariant of Eq.(\ref{Cayleygen}) negative or positive.

The two real SLOCC classes can alternatively be characterized by
the property that ${\cal B}_{AB}=\delta_{AB}$ or of the form ${\rm
diag}\{1,1,1,-1,-1,-1,-1\}$. In the first case one can calculate
the Hodge dual $\ast\phi$ of $\phi=\varrho +\omega\wedge e^7$ with
respect to the metric of Eq.(\ref{metrikahoz2}) \beq
\ast\phi=\hat{\varrho}\wedge e^7-\sigma,\qquad
\sigma=\frac{1}{2}\omega\wedge\omega, \eeq \noindent \beq
\hat{\varrho}=e^{456}-e^{234}+e^{135}-e^{126},\qquad \omega=
e^{14}+e^{25}+e^{36}. \eeq \noindent Notice that
a calculation shows
that  $\hat{\varrho}$ is the Freudenthal dual of
$\varrho=e^{123}-e^{156} +e^{246}-e^{345}$. According to the
formula \beq
\frac{1}{4}\varrho\wedge\hat{\varrho}=\frac{1}{6}\omega\wedge\omega\wedge\omega
\eeq \noindent hence using $\omega\wedge\varrho=0$ and the
invariance properties of $V(\Phi)$ an alternative formula for
Hitchins functional on $M_7$ is\cite{Hitchin,Dijkgraaf} \beq
V_7(\Phi)=\int_{M_7}\Phi\wedge\ast_{\Phi}\Phi. \eeq \noindent

Recall that the SLOCC classes are all of the canonical form
$\omega_i\wedge e^7+\varrho_{a}$ where $i=0,1,2,3$ refers to the Slater rank\cite{Schlie} of the $\omega$  (for $i=3$ we have a full rank symplectic form) and $a=0,1,2,3,4$
labeling the five entanglement classes for six fermions with six modes.
For the six mode case and its STU truncation the degenerate cases have the interpretation as small black holes\cite{Kalloshlinde}.
What is the physical interpretation of the degenerate cases of the seven mode case, i.e. the classes $f_1,\dots f_8$ of Table I?

\section{Conclusions}

In this paper we put forward the proposal to regard the invariants
underlying the Hitchin functionals as entanglement measures for
special entangled systems. In this picture the nondegenerate class
of stable forms corresponds to the class of genuine entangled
(GHZ-like) states. This idea makes it possible to generalize the
BHQC substantially. Unlike in conventional treatments of the
subject where entanglement measures were directly related to the
Bekenstein-Hawking entropy formulas we have shown that it is more
natural to connect them to {\it action functionals}. From such
functionals one can recover the usual correspondence with the
Bekenstein-Hawking entropy merely at the {\it semiclassical
level}. Furthermore since one loop calculations based on {\it
quantization} of such functionals are also capable of reproducing
results obtained by topological string techniques, via the OSV
conjecture this interpretation also hints that one can use the
BHQC beyond the semiclassical level. This approach also has the
advantage that it suggests that one does not have to assume the
underlying manifold to be furnished with a special holonomy
(Calabi-Yau, $G_2$ etc.) structure from the start. On the contrary
these structures are arising as critical points of functionals
coming from measures of entanglement.
 Identifying Hitchin's invariants with measures of entanglement
 also makes it possible to reconsider previous results
of the BHQC on the attractor mechanism as a distillation procedure
within a nice and unified framework. As a side result we connected
the notion of the Freudenthal dual to the one of almost complex and
generalized almost complex structures on $M$. These structures are
integrable precisely when the Freudenthal dual form (or state) is
closed. Finally as an application to Quantum Information we have
seen that Hitchin's functional for $7$ dimensional manifolds gives
rise to a natural measure of entanglement playing a basic role in
understanding the SLOCC classes of three fermionic states with
seven modes. We observed that the analogue of the GHZ class
provides a representative state (the calibration form) which via
the correlations in its reduced density matrices might serve as a
candidate for "seeing the octonions in the lab".

Notice that for six dimensional manifolds all of our functionals
were based on special PVs coming from Freudenthal systems of
simple cubic Jordan algebras. These Jordan algebras are the
complexifications of the cubic ones of Hermitian matrices with
real, complex and quaternionic entries. In Table II. we briefly
summarized the properties of the relevant PVs as related to
Freudenthal systems. In this paper we have not yet mentioned the
string theoretical background of the octonionic case. This case with the corresponding functional based on the quartic invariant of $E_7$ should be
connected to the important new development of generalized
exceptional geometry\cite{Chull,Mgrana}. In this field there are reformulations
of the N=2 supergravity backgrounds arising in Type II string
theory in terms of quantities transforming under the $E_{7(7)}$
U-duality group. This formalism combines the pure spinors of the
Neveu-Schwartz sector connected to the degrees of freedom of
generalized complex geometry with the Ramond-Ramond sector giving
rise to an extended  version of generalized geometry. It would be
instructive to connect our approach based on Freudenthal systems
to these results.

Let finally discuss some of the important conceptual issues we
have not yet investigated. Throughout this paper we called
entangled "states"  objects  like $\Gamma\in H^3(M,{\mathbb Z})$
and $\varrho\in H^3(M,{\mathbb R})$ or $\varphi\in
H^{\bullet}(M,{\mathbb R})$ (where the latter is a polyform of
either even or odd degree). In particular we called the
representatives of cohomology classes  of the $M=T^2\times
T^2\times T^2$ STU case as "3-qubit states". Is there a physical
basis for calling such constructs "entangled states" of some kind?

First of all let us notice that the spaces of {\it real}
cohomology classes that show up in the Hitchin and generalized
Hitchin functionals are all {\it phase spaces} in the conventional
sense. The symplectic form is the usual one defined for
Freudenthal systems which is just the Mukai pairing for polyforms.
The Hamiltonians on these phase spaces are the functionals
themselves, the Freudenthal duals are the corresponding
Hamiltonian vector fields. Thanks to these properties in all cases
of the PVs of Table II. we can regard the elements of such
Freudenthal systems as "classical states".

On the other hand our classical phase spaces are locally the
moduli spaces of complex\cite{Hitchin}, generalized
complex\cite{genHitchin}, and probably generalized exceptional
structures. However these spaces are in turn also complex ones so
we should see a complex structure on them. As a byproduct of this
observation beyond the classical one there should be extra
structures playing an important role. In the case of the Hitchin
functional we can illustrate this as follows.

{\bf 1.} One can embed the real cohomology classes into
$H^3(M,{\mathbb C})$. This corresponds to the fact that the stable
open orbit can be given a structure of a pseudo K\"ahler
manifold\cite{Hitchin} with signature $(1,h^{2,1})$ with the
complex structure defined by the derivative of the map that
associates to a state its Freudenthal dual. This complex structure
is acting on $H^3(M,{\mathbb C})$ as $+i$ on $H^{3,0}\oplus
H^{2,1}$ and as $-i$ on $H^{0,3}\oplus H^{1,2}$.

 {\bf 2.} One can also embed the real cohomology
classes into the space of complex ones furnished with a Hermitian
inner product of Eq.(\ref{inner}). The rationale for doing this is
encoded into the expansion of $[\Gamma]=\varrho$ in the Hodge
diagonal basis (e.g. like the expansion of Eq.(\ref{expand})).
Notice that the Hodge star is acting on $H^3(M,{\mathbb R})$ as
$+i$ on $H^{3,0}\oplus H^{1,2}$ and as $-i$ on $H^{0,3}\oplus
H^{2,1}$ (in the STU case $\ast$ is just $i$ times the parity
check operator i.e. $i\sigma_3\otimes\sigma_3\otimes \sigma_3$).
This defines an alternative complex structure and embedding for
$H^3(M,{\mathbb R})$. Notice also that in this case
$\vert\vert\Gamma\vert\vert^2$ is positive and related to the
Black Hole Potential. Since the Hodge diagonal basis is depending
on the coordinates $\tau,\overline{\tau}$ of the moduli space
${\cal M}$ of $M$ we obtain "states" with complex amplitudes
depending on the charges and the moduli. This is the setting which
made it possible to regard our real states as also elements of a
complex finite dimensional Hilbert space making the entanglement
interpretation useful.

Do not confuse our entangled "states" with the ones discussed in
topological string theory. The two different states are related by
geometric quantization.

First recall the physical meaning of case {\bf 1}. According to
the OSV conjecture the partition function for BPS black holes in
Calabi-Yau compactifications of type II string theory is equal to
the product of partition sums of topological strings. The
topological string partition function can also be
interpreted\cite{Witten} as a wave function obtained by quantizing
our classical phase space $H^3(M,{\mathbb R})$ . The idea is that
there should be a state $\vert\Psi\rangle$ which contains the
background independent information of topological string theory.
In order to carry out this (geometric) quantization a polarization
is needed. The polarization which is used for this quantization is
the one of {\bf 1.} and again depending on the coordinates
$\tau,\overline{\tau}$ of the moduli space ${\cal M}$ of $M$. The
dependence on these coordinates is expressed in the holomorphic
anomaly equation\cite{Cecotti,Verlinde}. As we know\cite{Neitzke}
the Hermitian metric constructed from the canonical symplectic
structure on $H^3(M,{\cal R})$ and this complex structure is not
positive definite, but rather of signature $(1,h^{2,1})$. The
quantization is carried out by elevating the expansion
coefficients of Eq.(\ref{expand}) (i.e. the amplitudes of our entangled
states) to moduli-dependent annihilation operators\cite{Neitzke}
and then constructing coherent states. This results in
non-normalizable states. However in this approach the holomorphic
dependence of the complex structure on $\tau$ is manifest.

On the other hand using {\bf 2.} the Weyl
polarization\cite{Neitzke} provided by the Hodge star we have a
positive definite metric, however the holomorphic dependence of
the complex structure is lost. This polarization is not suitable
for studying the holomorphic anomaly equations, however directly
connected to our entanglement interpretation. Moreover, it is
probably more natural for finding its role in the non-BPS version
of the OSV conjecture\cite{Saraikin} where the holomorphic
structure is lost. Can we relate somehow this non-BPS branch to the
real orbit with ${\cal D}>0$  of Hitchin's functional?


\begin{table}[h]
\centering
\begin{tabular}{|c|c|c|c|}
\hline
$\mathfrak{J}$ & Inv$(\mathfrak{M})$ & dim$\mathfrak{M}$ & Hitchin functional\\
\hline \hline
$\mathcal{H}_3(\mathbb{R})$ & $Sp(6,\mathbb{C})$ & 14 & Constrained Hitchin \\
$\mathcal{H}_3(\mathbb{C})$ & $SL(6,\mathbb{C})$ & 20 & Hitchin \\
$\mathcal{H}_3(\mathbb{H})$ & $Spin(12,\mathbb{C})$ & 32 & Generalized Hitchin\\
$\mathcal{H}_3(\mathbb{O})$ & $E_{7}(\mathbb{C})$ & 56 & Generalized Exceptional\\ \hline
\end{tabular}
\caption{Freudenthal triple systems ($\mathfrak{M(J)}$) over cubic Jordan algebras ($\mathfrak{J}$), their automorphism groups (Inv$\mathfrak{M(J)}$) and the corresponding Hitchin functional.}
\label{tab:2}
\end{table}


\section{Acknowledgement}
This work was supported by the New Hungary Development Plan
(Project ID: T\'AMOP-4.2.1/B-09/1/KMR-2010-002).

\section{Appendix}

In this Appendix we would like to establish a dictionary for the generalized Hitchin functional between the languages based on polyforms and the Freudenthal
systems based on the Jordan algebra of quaternion Hermitian $3\times 3$ matrices.
Let $W$ be a six dimensional {\it real} vector space and $W^{\ast}$ its dual.
The basis vectors for these spaces will be denoted by $\{e_i\}$ and $\{e^i\}$
 $i=1,2\dots 6$ respectively.
 There is a natural symmetric bilinear form on the space $W\oplus W^{\ast}$
 given by
\beq
(v+\omega,u+\sigma)=\frac{1}{2}(\omega(u)+\sigma(v)),\qquad
v,u\in W,\quad \omega,\sigma\in W^{\ast}.
\eeq
\noindent
This symmetric form has signature $(6,6)$ and defines the non-compact orthogonal group $O(W\oplus W^{\ast})\simeq O(6,6)$.
By noticing that
\beq
\wedge^{12}(W\oplus W^{\ast})=\wedge^6W\otimes \wedge^6 W^{\ast}
\eeq
\noindent
and using the natural pairing between the latter two one can define a canonical orientation. The group preserving the symmetric form taken together with this orientation is $SO(W\oplus W^{\ast})\simeq SO(6,6)$.
The Lie algebra of this group is defined as usual by
\beq
so(W\oplus W^{\ast})=\{T\vert (Tu,v) +(u, Tv)=0,\quad
u,v\in W\oplus W^{\ast}\}
\eeq
\noindent
and can be parametrized as
\beq
T=\begin{pmatrix}A&\beta\\B&-A^{\ast}\end{pmatrix}.
\eeq
\noindent
Here
\beq
A\in {\rm End}(W), \qquad A={A^i}_je^j\otimes e_i,
\label{A}
\eeq
\noindent
\beq
B\in \Lambda^2W^{\ast}: W\to W^{\ast},\qquad B=\frac{1}{2}B_{ij}e^i\wedge e^j,
\label{B}
\eeq
\noindent
\beq
\beta\in \Lambda^2W: W^{\ast}\to W,\qquad \beta=\frac{1}{2}\beta^{ij}e_i\wedge e_j.
\label{beta}
\eeq
\noindent
This shows that $so(W\oplus W^{\ast})=\Lambda^2(W\oplus W^{\ast})=End(W)\oplus
\Lambda^2W^{\ast}\oplus \Lambda^2W$.

Let us now define the Clifford algebra ${\rm Cliff}(W\oplus W^{\ast})$
by the relation
\beq
w^2=(w,w){\bf 1}, \forall w\in W\oplus W^{\ast}.
\eeq
\noindent
The Clifford algebra can be represented on the space $\wedge^{\bullet} W^{\ast}$
of polyforms by
\beq
(v+\omega)\cdot \varphi=i_v\varphi+\omega\wedge\varphi,\qquad \varphi\in \wedge^{\bullet} W^{\ast}.
\eeq
\noindent
Indeed,
\beq
(v+\omega)^2\cdot \varphi=i_v(\omega\wedge\varphi)+\omega\wedge (i_v\varphi)=(i_v\omega)\varphi=\langle v+\omega,v+\omega\rangle
\eeq
\noindent
hence we have an algebra representation.
This formula also gives rise to the standard spin representation
hence the exterior algebra provides a natural description of spinors provided\cite{Chevalley} we tensor with the one-dimensional space $(\wedge^6W)^{1/2}$.
Hence the representation space is
\beq
S=\wedge^{\bullet}W^{\ast}\otimes(\wedge^6W)^{1/2}.
\label{spinorspace}
\eeq
\noindent
We can decompose the space of spinors to positive and negative chirality elements $S=S^+\oplus S^-$ under the $\pm 1$ eigenspaces of the volume element of the Clifford algebra. These are simply exterior forms of even and odd degree
\beq
S^+=\wedge^{ev}W^{\ast}\otimes(\wedge^6W)^{1/2},\quad
S^-=\wedge^{odd}W^{\ast}\otimes(\wedge^6W)^{1/2}.
\label{chiral}
\eeq
\noindent
They are irreducible under the double cover of $SO(W\oplus W^{\ast})$ the spin group ${\rm Spin}(W\oplus W^{\ast})$ consisting of products with an {\it even} number of elements $w_1w_2\dots w_{2r}$, where $(w_i,w_i)=\pm 1$.

Since $so(W\oplus W^{\ast})$ can also be embedded
in the Clifford algebra one can calculate the spinorial action
of $A,B$ and $\beta$ of Eqs.(\ref{A})-(\ref{beta}) on $\wedge^{\bullet}W^{\ast}$.
One can then show that\cite{Gualtieri}
the spinorial versions of $A,B$ and $\beta$ are respectively
$\frac{1}{2}{A^i}_j(e_ie^j-e^je_i)$, $\frac{1}{2}B_{ij}e^je^i$ and
$\frac{1}{2}\beta^{ij}e_je_i$.
As a result
the spinorial actions take the form
\beq
A\cdot\varphi=\frac{1}{2}{\rm Tr}A-A^{\ast}\varphi=\frac{1}{2}{\rm Tr}A-{A^i}_j
e^j\wedge i_{e_i}\varphi,
\label{As}
\eeq
\noindent

\beq
B\cdot\varphi=-B\wedge \varphi=\frac{1}{2}B_{ij}e^j\wedge (e^i\wedge \varphi)
\label{Bs}
\eeq
\noindent

\beq
\beta\cdot\varphi=i_\beta\varphi=\frac{1}{2}\beta^{ij}i_{e_j}(i_{e_i})\varphi
\label{betas}
\eeq
\noindent
An important corollary of Eq.(\ref{As}) is that after exponentiation the spinorial action of an element  $L\in GL^+(W)$
can be expressed as

\beq
L\cdot \varphi =\sqrt{{\rm Det} L}(L^{\ast})^{-1}\varphi
\label{density}
\eeq
\noindent
giving a rationale for the appearance of the factor $(\wedge^6W)^{1/2}$ in
Eq. (\ref{spinorspace}).

Let us now complexify our $W$ to $V=W\otimes {\mathbb C}={\mathbb C}^6$ and
let $\varphi\in \wedge^{ev} V^{\ast}$  be a polyform
of even degree.
Then we have
\beq
\varphi=\varphi_{0}+\varphi_{2}+\varphi_{4}+\varphi_{6},\qquad \varphi_{p}\in   \wedge^{p}V^{\ast}.
\eeq

Since $B\cdot\varphi=-B\wedge \varphi$, for a $B=\frac{1}{2}B_{ij}e^i\wedge e^j\in \wedge^2V^{\ast}$ we have the spinorial action of $e^{-B}$ on the special form $\varphi_0\equiv 1$ as

\begin{eqnarray}
e^{-B}\cdot 1&=&(1+B+ \frac{1}{2}B\wedge B+\frac{1}{6}B\wedge B\wedge B)\cdot 1 \\ &=&
1+\sum_{i<j}B_{ij}e^i\wedge e^j+\sum_{i<j}{\rm Pf}(B_{(ij)} )\ast(e^i\wedge e^j)
+{\rm Pf}(B){\epsilon} .\nonumber
\end{eqnarray}

Here ${\epsilon}=e^1\wedge e^2\wedge e^3\wedge e^4\wedge e^5\wedge e^6$ and the Pfaffian of the $6\times 6$ {\it complex} matrix $B_{ij}$ is

\beq
{\rm Pf}(B)=\frac{1}{3!2^3}\varepsilon^{ijklmn}B_{ij}B_{kl}B_{mn}.
\eeq
\noindent
On the other hand ${\rm Pf}(B_{(ij)})$ is the Pfaffian of the $4\times 4$ matrix obtained from the original $6\times 6$ one after omitting the $(i,j)$th rows and columns.
Hence for example
\beq
{\rm Pf}(B_{(56)})=B_{12}B_{34}-B_{13}B_{24}+B_{14}B_{23},\qquad \ast(e^5\wedge e^6)=e^1\wedge e^2\wedge e^3\wedge e^4.
\eeq
\noindent

Let us now recall that ${\rm Skew}(6,{\mathbb C})\simeq {\rm Herm}(3,{\mathbb H})\otimes {\mathbb C}$, i.e. the space of $6\times 6$ skew-symmetric matrices
with complex entries can be identified with the cubic Jordan algebra of quaternion Hermitian matrices when the quaternions are replaced by biquaternions.
An identification of these objects is given as follows
\beq
{\cal B}=\begin{pmatrix}\alpha&c&\overline{b}\\\overline{c}&\beta&a\\b&\overline{a}&\gamma\end{pmatrix}\leftrightarrow
B=\begin{pmatrix}\alpha\epsilon&c\epsilon&\tilde{b}\epsilon\\\tilde{c}\epsilon&
\beta\epsilon&a\epsilon\\b\epsilon&\tilde{a}\epsilon&\gamma\epsilon\end{pmatrix}.
\label{identifi}
\eeq
Here on the left hand side $\alpha,\beta,\gamma\in{\mathbb C}$,
$a,b,c\in {\mathbb H}\otimes {\mathbb C}$, and overline refers to quaternionic conjugation.
On the right hand side we have $\alpha,\beta, \gamma\in {\mathbb C}$,
$a,b,c\in Matr(2,{\mathbb C})$ i.e. $2\times 2$ complex matrices,               $\tilde{a}\equiv -\epsilon a^T\epsilon$ with $\epsilon$ the standard $SL(2,{\mathbb C})$ invariant antisymmetric $2\times 2$ matrix with $\epsilon_{12}=1$.
One can check that $B^T=-B$.

Now in the language of cubic Jordan algebras the cubic norm $N({\cal B})={\rm Det}({\cal B})$  is the determinant of the $3\times 3$ matrix with biquaternionic entries.
It can be checked that it corresponds to the Pfaffian of the $6\times 6$ antisymmetric matrix with complex entries, i.e.
\beq
{\rm Det}({\cal B})\leftrightarrow {\rm Pf}(B).
\eeq
Moreover, for elements of $Herm(3,{\mathbb H})\otimes {\mathbb C}$ one can define the quadratic {\it sharp} map by
\beq
{\cal B}\mapsto {\cal B}^{\sharp}={\cal B}^2-{\rm Tr}({\cal B}){\cal B}
+\frac{1}{2}(({\rm Tr}({\cal B}))^2-{\rm Tr}({\cal B}^2))I
,
\label{sharp}
\eeq
satisfying ${\cal B}{\cal B}^{\sharp}={\rm Det}({\cal B})I$ with $I$ the $3\times 3$ unit matrix.
The polarization of the sharp map is
\beq
{\cal B}\times{\cal C}=({\cal B}+{\cal C})^{\sharp}-{\cal B}^{\sharp}-{\cal C}^{\sharp}.
\label{pol}
\eeq
\noindent
Now one can check that
\beq
{\cal B}^{\sharp}\leftrightarrow {\rm Pf}(B_{(\cdot\cdot)} ).
\eeq

As a result of these considerations one can have the correspondence
\beq
e^{-B}\cdot 1\leftrightarrow (1,{\cal B}, {\cal B}^{\sharp},{\rm Det}({\cal B}))\in{\mathbb C}\oplus {\cal J}\oplus {\cal J}\oplus {\mathbb C},
\label{egyhat}
\eeq
\noindent
where we denoted the cubic Jordan algebra ${\rm Herm}(3,{\mathbb H})\otimes {\mathbb C}$ by ${\cal J}$.
The algebraic object ${\mathbb C}\oplus {\cal J}\oplus {\cal J}\oplus {\mathbb C}$ is called the Freudenthal triple system ${\cal F}({\cal J})$  associated to the cubic  Jordan algebra ${\cal J}$.
In particular one can see that $e^{-B}\cdot 1$ can be mapped to a special element of ${\cal F}({\cal J})$.
Now it is straightforward to elaborate the whole correspondence between
the action of $Spin(12,{\mathbb C})$ on the space of spinors
$S=S^+=\Lambda^{ev}V^{\ast}\otimes (\Lambda^6 V)^{1/2}$  and the {\it conformal group} of ${\cal J}$, ${\rm Conf}({\cal J})$, acting on ${\cal F}$.

The conformal group of ${\cal J}$ is the group of rational transformations
of ${\cal J}$ generated by
the translations $({\cal T})$, inversions (${\cal I}$), and transformations (${\cal L}$) belonging to the structure group of ${\cal J}$ (linear bijections of ${\cal J}$ leaving invariant the norm $N$ up to a character ${\chi}$). The translations and inversions  are of the following form
\beq
{\cal T}_{{\cal B}}:{\cal Z}\mapsto {\cal Z}+{\cal B},
\eeq
\noindent
\beq
{\cal I}:{\cal Z}\mapsto -{\cal Z}^{-1}.
\eeq
\noindent

It is known\cite{Gindikin} that there is a projective irreducible representation of ${\rm Conf}({\cal J})$ on ${\cal F}({\cal J})$ which is of the form

\beq
{\pi}(g)(\eta, y,x,\xi)=(\eta^{\prime},y^{\prime},x^{\prime},\xi^{\prime})\in {\cal F}, \quad g\in {\rm Conf}({\cal J})
\eeq
\noindent
with the translations $\pi({\cal T}_{-{\cal B}})$ act as
\begin{eqnarray}
\eta^{\prime}&=&\eta\\
y^{\prime}&=&y+\eta {\cal B}\\
x^{\prime}&=&x+{\cal B}\times y+\eta {\cal B}^{\sharp}\\
\xi^{\prime}&=&\xi+{\rm Tr}({\cal B}x)+{\rm Tr}({\cal B}^{\sharp}y)+\eta{\rm Det}{\cal B},
\label{trans}
\end{eqnarray}
(for the definition of ${\cal B}\times y$ see Eq.(\ref{pol})). For the inversions $\pi({\cal I})$ we have
\beq
\eta^{\prime}=\xi,\quad y^{\prime}=-x,\quad x^{\prime}=y,\quad \xi^{\prime}=-\eta,
\eeq
\noindent
and finally for $\pi({\cal L})$ one gets
\begin{eqnarray}
\label{Afreud}
\eta^{\prime}&=&\chi({\cal L})^{-1/2}\eta\\\nonumber
y^{\prime}&=&{\chi}({\cal L})^{-1/2}{\cal L}(y)\\\nonumber
x^{\prime}&=&{\chi}({\cal L})^{1/2}{\cal L}^{\ast-1}(x)\\\nonumber
\xi^{\prime}&=&\chi({\cal L})^{1/2}\xi.
\end{eqnarray}
\noindent

By virtue of Eqs.(\ref{egyhat}) and (\ref{trans}) we have
\beq
e^{-B}\cdot 1\leftrightarrow {\pi}({\cal T}_{-{\cal B}})(1,0,0,0)=(1,{\cal B},{\cal B}^{\sharp},{\rm Det}{\cal B}).
\label{naezaz}
\eeq
\noindent

Now by associating a polyform to an element of ${\cal F}$ as

\beq
(\varphi_0,\varphi_2,\varphi_4,\varphi_6)\leftrightarrow (\eta,y,x,\xi)
\label{corralap}
\eeq
\noindent
one can check that
\beq
e^{-B}\cdot \varphi\leftrightarrow \pi({\cal T}_{-{\cal B}})(\eta,y,x,\xi).
\eeq
\noindent
Similarly recalling Eqs. (\ref{density}) and (\ref{Afreud}) for the $L\in GL(6,{\mathbb C})$ action
we get the correspondence
\beq
L\cdot\varphi\leftrightarrow \pi({\cal L})(\eta,y,x,\xi),
\eeq
\noindent
with the character $\chi({\cal L})\leftrightarrow ({\rm Det} L)^{-1}$.
Finally the correspondence for the $\beta$ transform takes the form
\beq
e^{\beta}\cdot\varphi\leftrightarrow \pi({\cal I}^{-1}\circ {\cal T}_{\cal B}\circ {\cal I})(\eta,y,x,\xi).
\eeq
\noindent

The upshot of these considerations is that we managed to represent the $Spin(12,{\mathbb C})$ action on polyforms of even degree as the action of
${\rm Conf}({\cal J})$ on the Freudenthal triple system ${\cal F}({\cal J})$.
This construction enables the identification of the generalized Hitchin functional with the quartic invariant for ${\cal F}({\cal J})$.

In order to do this, recall that for ${\cal F}({\cal J})$
we can define a symplectic form and a quartic polynomial, both invariant
under ${\rm Conf}({\cal J})$.
The symplectic form can easily be related to the
symplectic form of Hitchin\cite{genHitchin}.
The latter is defined as
\beq
\langle\varphi,\psi\rangle=\varphi_0\psi_6-\varphi_2\psi_4+\varphi_4\psi_2-\varphi_6\psi_0\in \wedge^6V^{\ast}\otimes((\wedge^6V)^{1/2})^2={\mathbb C}.
\label{symphitchin}
\eeq
\noindent
On the other hand the symplectic form on ${\cal F}({\cal J})$ takes the form
\beq
\{p,p^{\prime}\}=\eta\xi^{\prime}-{\rm Tr}(y\bullet x^{\prime})+{\rm Tr}(x\bullet y^{\prime})-\xi\eta^{\prime},\qquad p=(\eta,y,x,\xi),\quad p^{\prime}=(\eta^{\prime},y^{\prime},x^{\prime},\xi^{\prime}).
\label{sympfreud}
\eeq
\noindent
Here $\eta,\xi\in {\mathbb C}$ and $x,y\in {\rm Herm}(3,{\mathbb H})\otimes {\mathbb C}$, $x\bullet y=\frac{1}{2}(xy+yx)$ is the Jordan product.
Clearly, by virtue of the correspondence Eq.(\ref{corralap}) and the identity
$\varphi_2\psi_4\leftrightarrow {\rm Tr}(y\bullet x^{\prime})$  these structures are mapped to each other.

The quartic invariant for ${\cal F}({\cal J})$ takes the following form\cite{Krutelevich}
\beq
q(p)=-[\eta\xi-{\rm Tr}(x\bullet y)]^2+4{\rm Tr}(x^{\sharp}\bullet y^{\sharp})-4\eta{\rm Det}(x)-4\xi{\rm Det}(y),\qquad p=(\eta,y,x,\xi).
\label{quarticinv}
\eeq
\noindent
By virtue of the identification in Eq.(\ref{identifi}) an alternative formula can also be given
\beq
q(p)=-[\eta\xi-\sum_{i<j}x_{ij}y_{ij}]^2+4\sum_{i<j}{\rm Pf}(x_{(ij)})
{\rm Pf}(y_{(ij)})-4\eta{\rm Pf}(x)-4\xi{\rm Pf}(y)
\label{quart2}
\eeq
\noindent
where $\eta,\xi\in {\mathbb C}$ and $x,y$ are $6\times 6$ skew-symmetric matrices
with complex entries.
The last version of the quartic invariant can easily be related to the coefficients of the polyforms $(\varphi_0,\varphi_2,\varphi_4,\varphi_6)$ needed for the
explicit expression of the generalized Hitchin functional.
For this we just have to parametrize these component forms as

\begin{eqnarray}
\varphi_0&=&\eta \otimes ({\epsilon}^{\ast})^{1/2}\\\nonumber
\varphi_2&=&\frac{1}{2!}y_{ij}e^i\wedge e^j\otimes ({\epsilon}^{\ast})^{1/2} \\\nonumber
\varphi_4&=&\frac{1}{4!}\frac{1}{2!}x_{ij}{\varepsilon^{ij}}_{klmn}e^k\wedge e^l\wedge e^m\wedge e^n\otimes ({\epsilon}^{\ast})^{1/2}\\\nonumber
\varphi_6&=&\xi {\epsilon}\otimes
({\epsilon}^{\ast})^{1/2}
\label{masod}
\end{eqnarray}
\noindent
where  ${\epsilon}=e^1\wedge e^2\wedge\dots \wedge e^6$ and ${\epsilon}^{\ast}\equiv e_1\wedge e_2\wedge\dots \wedge e_6$.

\end{document}